\documentclass[aps,pra,twocolumn,floatfix,reprint]{revtex4}
\usepackage{amsfonts}
\usepackage{mathrsfs}
\usepackage{amsmath}
\usepackage{color}
\usepackage{graphicx}
\usepackage{bm}
\usepackage{amssymb}
\usepackage{xspace}
\usepackage{epstopdf}
\usepackage{dcolumn}
\usepackage{tabularx}
\usepackage{longtable}
\usepackage{footnote}
\usepackage[colorlinks=true, pdfstartview=FitV, linkcolor=blue, citecolor=blue, urlcolor=blue]{hyperref}
\usepackage[normalem]{ulem}
\usepackage{verbatim}
\begin{document}

\title{Magnetic solitons in Rabi-coupled Bose-Einstein condensates}
\author{Chunlei Qu$^{1}$}
\email{chunleiqu@gmail.com}
\author{Marek Tylutki$^{1}$}
\author{Sandro Stringari$^{1}$}
\author{Lev P. Pitaevskii$^{1,2}$}

\affiliation{
$^{1}$INO-CNR BEC Center and Dipartimento di Fisica, Universit\`a di Trento, 38123 Povo, Italy \\
$^{2}$Kapitza Institute for Physical Problems RAS, Kosygina 2, 119334 Moscow, Russia}

\begin{abstract}
We study magnetic solitons, solitary waves of spin polarization (i.e.,
magnetization), in binary Bose-Einstein condensates in the presence of Rabi
coupling. We show that the system exhibits two types of magnetic solitons,
called $2\pi$ and $0\pi$ solitons, characterized by a different behavior of
the relative phase between the two spin components. $2\pi$ solitons exhibit
a $2\pi$ jump of the relative phase, independent of their velocity, the
static domain wall explored by Son and Stephanov being an example of such $%
2\pi$ solitons with vanishing velocity and magnetization. $0\pi$ solitons
instead do not exhibit any asymptotic jump in the relative phase. Systematic
results are provided for both types of solitons in uniform matter. Numerical
calculations in the presence of a one-dimensional harmonic trap reveal that
a $2\pi$ soliton evolves in time into a $0\pi$ soliton, and vice versa,
oscillating around the center of the trap. Results for the effective mass,
the Landau critical velocity, and the role of the transverse confinement are
also discussed.
\end{abstract}

\maketitle

\section{Introduction}
\label{sec:intro}

Solitary waves are nontrivial collective excitations that appear in a
wide variety of systems in different physical branches including classical
fluids, cosmology~\cite{Kibble1976}, condensed matter~\cite{Emori2013,Su1979}%
, optics~\cite{Mollenauer1980}, and cold atoms~\cite{Burger1999,Denschlag2000}.
Despite the fact that they do not correspond to the ground states of the
systems, these solitary waves can be stable and live for a long time
under certain physical conditions, which may have important applications for
information processing. Because of the tunability of the interaction
coupling constants and the absence of disorder, ultracold atomic gases
provide an ideal playground for the observation of these excitations. Since
the first realization of Bose-Einstein condensate with alkali atoms, various
solitary waves and other quantum defects have been experimentally observed and/or theoretically
investigated, such as scalar solitons~\cite%
{Burger1999,Denschlag2000,Khaykovich2002,Lamporesi2013}, vector solitons~\cite%
{Busch2001,Becker2008,Hamner2011,Liu2009}, domain walls~\cite{Hall1998,Trippenbach2000,Son2002},
vortices~\cite{Matthews1999a,Madison1999,AboShaeer2001}, and skyrmions~\cite%
{Khawaja2001,Leslie2009}.

The application of a coherent coupling between two internal states is a
powerful tool for the control of spinor condensates with external fields~%
\cite{Matthews1999b,Zibold2010}. In this work, we consider a
two-component Bose-Einstein condensate in the presence of a weak Rabi coupling,
\begin{equation}
\Omega\ll \mu/\hbar, \label{eq:conditionOmega}  
\end{equation}
where $\mu$ is the chemical potential of the system, the corresponding
solitons being intrinsically different from those in the absence
of Rabi coupling~\cite{Hamner2013,Congy2016,Gallemi2016}. Useful simplifications in the
determination of the solitonic solutions in uniform matter take place when
the intraspecies coupling constants are equal ($g\equiv g_{11}=g_{22}$) and
very close to the inter-species coupling $g_{12}$, i.e., 
\begin{equation}
\delta g\equiv g-g_{12}\ll g,  \label{eq:condition1}
\end{equation}%
with $\delta g>0$ in order to ensure miscibility even in the absence of Rabi
coupling~\cite{footnote0}. Conditions (\ref{eq:conditionOmega}) and (\ref{eq:condition1}) ensure that
the total density $n=n_{1}+n_{2}$ is only weakly affected by the presence of the
soliton and can be considered a constant $n=\mu/g$, thereby reducing the relevant variables
of the problem to the spin
density $n_{1}-n_{2}$ and to the phases of the two spin components (see Sec.~\ref{sec:SSsolution} for a discussion of the accuracy of the constant-total-density approximation). For this
reason the corresponding solutions are called magnetic solitons.
Condition (\ref{eq:condition1}) is fulfilled, for example, by the $|F=1;
m_F=\pm1 \rangle$ hyperfine states of $^{23}$Na.

\begin{figure}[tbp]
\includegraphics[width=6cm]{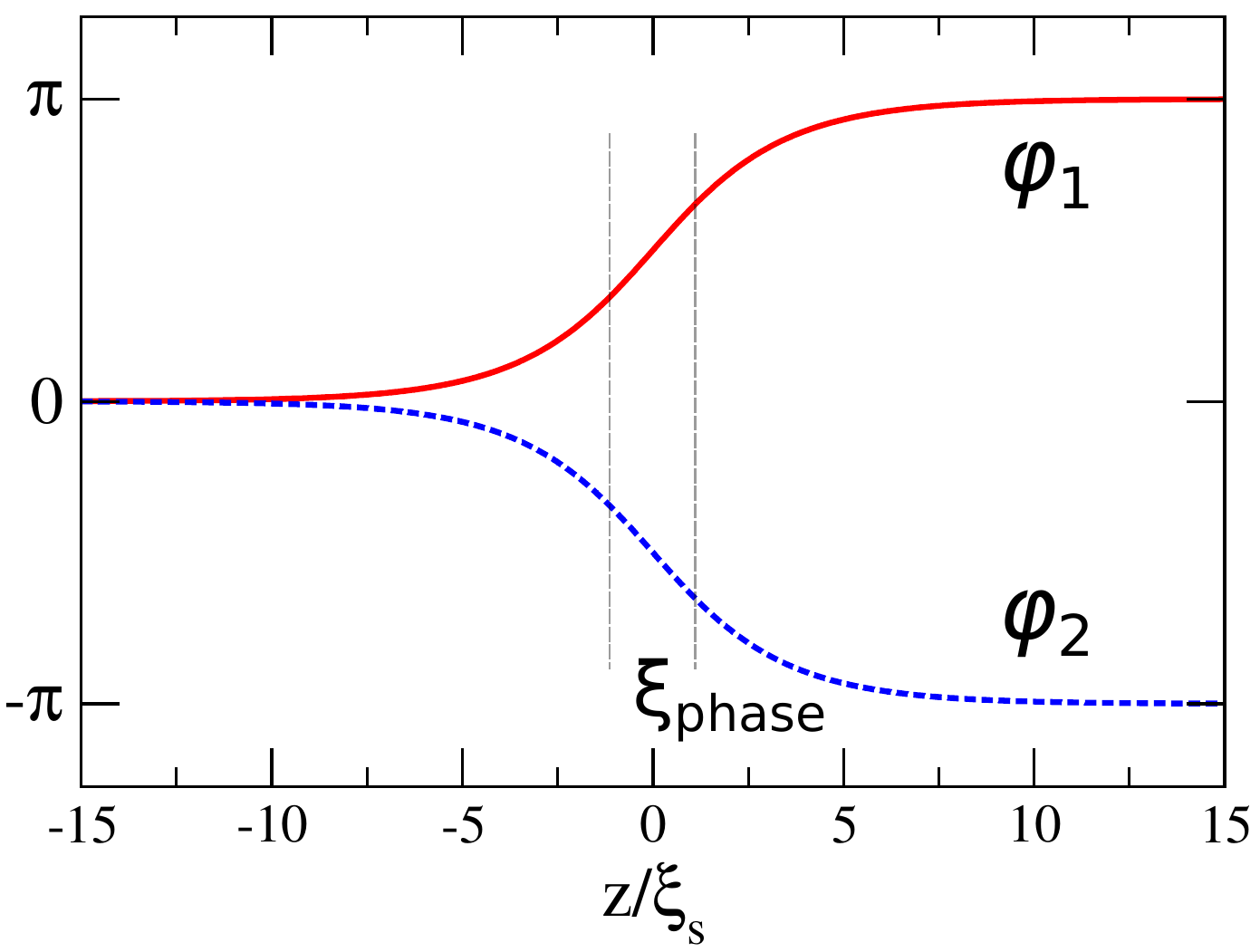}
\caption{Phase structure of the static Son-Stephanov domain wall. The
relative phase $\varphi_A=\varphi_1-\varphi_2$ of the two spin
states exhibits a $2\protect\pi$ jump when
one moves from $-\infty$ to $+\infty$. The width of the domain wall is fixed
by the characteristic length of the relative phase $\protect\xi_{\text{phase}%
}=\protect\sqrt{\hbar/2m\Omega}$, where $\Omega$ is the Rabi coupling.}
\label{fig:illustrate}
\end{figure}

We note that magnetic solitons have been predicted in the absence
of Rabi-coupling where the relative phase of the two components exhibits a $%
\pi$ phase jump across the soliton~\cite{Qu2016} (see also Ref.~\cite%
{Congy2016b} for more general solutions available under the same condition,~(%
\ref{eq:condition1}), and Ref.~\cite{Engels2016} for a recent experimental observation). In the presence of Rabi coupling, the relative phase $%
\varphi_A$ should satisfy the condition $\cos \varphi_A=1$ at large
distances from the soliton, which implies that the jump of the relative
phase must be equal to $2n\pi$ with $n=0,\pm 1,\ldots$.

A prominent example of a solitonic solution in a Rabi-coupled binary
condensate is the static domain wall identified by Son and Stephanov in 2002~%
\cite{Son2002} by considering two equally populated spin states coupled by a
weak Rabi coupling of strength [characterized by Eq.~(\ref{eq:conditionOmega})] in uniform matter. Under
assumption (\ref{eq:condition1}) these authors found a metastable solution,
corresponding to a local minimum of the energy functional, characterized by
the $2\pi$ jump of the relative phase of the two components across the wall
(see Fig.~\ref{fig:illustrate}). This static soliton is characterized by the
absence of magnetization (i.e., the spin is balanced) and corresponds to a metastable solution of the
coupled Gross-Pitaevskii equations (GPEs) if the condition 
\begin{equation}
\hbar \Omega <\hbar\Omega_\text{c}\equiv \frac{1}{3}n \delta g,
\label{eq:condition2}
\end{equation}%
is satisfied. For larger values of $\Omega$ the static domain wall
does not correspond to a local minimum of the energy functional and the resulting
configuration is consequently unstable in uniform matter~\cite{Son2002} (see also Ref.~\cite{Usui2015}).
Actually the magnetization of the domain wall becomes energetically profitable. In this paper we assume condition (\ref{eq:conditionOmega}) and mainly focus
on configurations which satisfy the stability condition, (\ref{eq:condition2}).

The absence of magnetization of the static domain wall makes its
experimental detection difficult. In this work we show that the
Son-Stephanov domain wall exhibits a magnetization when it moves, thereby
opening realistic perspectives for its experimental detection.

\begin{figure}[tbp]
\includegraphics[width=8cm]{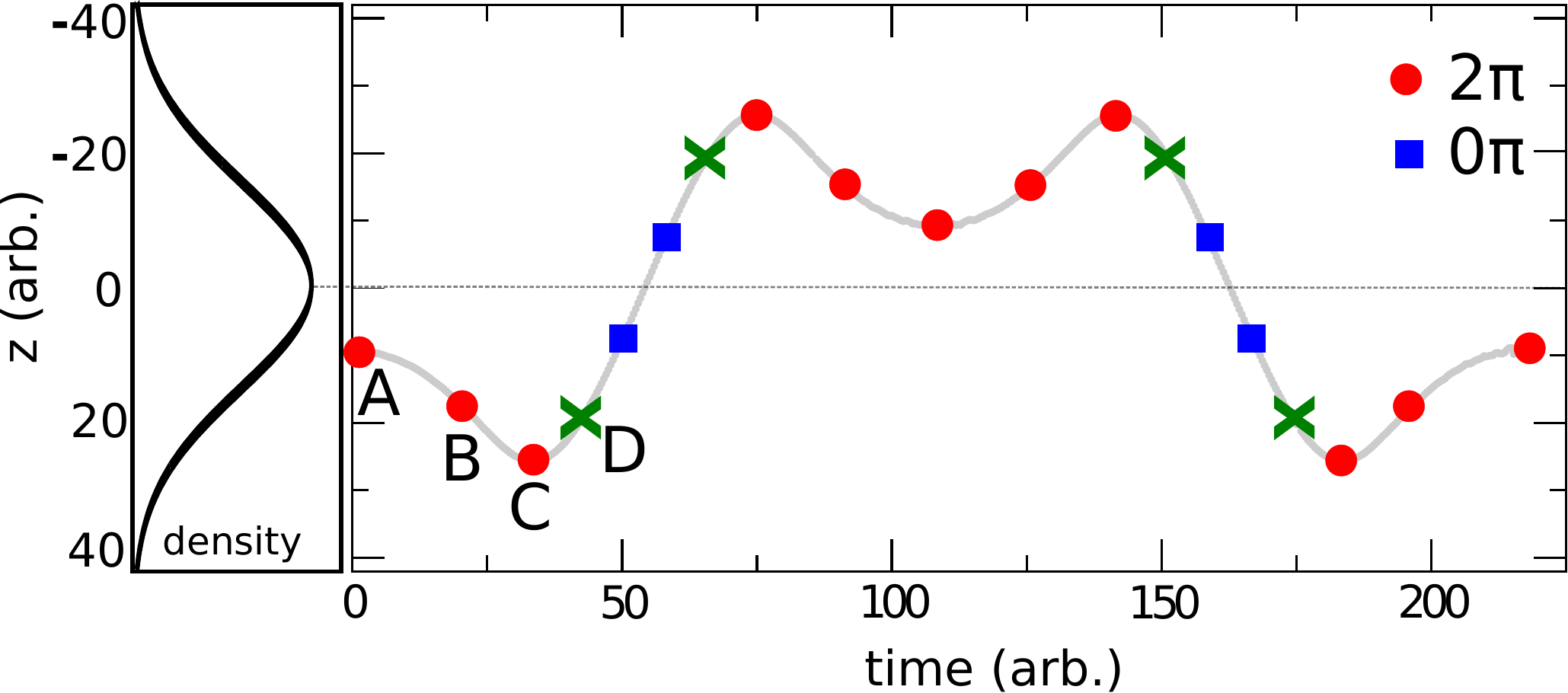}
\caption{Illustration of the dynamics of magnetic solitons in a harmonic
trap. At $t=0$, a static Son-Stephanov domain wall, characterized by a $2%
\protect\pi$ jump in the relative phase, is imprinted at the right (point $A$%
) of the trap center . The $2\protect\pi$ soliton starts moving towards the
periphery, and soon after the reflection, it evolves into a $0\protect\pi$
soliton (point $D$) (see text). $2\protect\pi$ solitons are indicated by
red circles; $0\protect\pi$ solitons, by blue squares.
Green X's are points indicating the transformation between $2\protect%
\pi$ and $0\protect\pi$ solitons.}
\label{fig:trapevolution}
\end{figure}

In order to generate moving magnetic solitons we found it convenient to
imprint the phase of the static Son-Stephanov domain wall (see Fig.~\ref%
{fig:illustrate}), with its center displaced from the center of the trap
(see Fig.~\ref{fig:trapevolution}), and to follow the numerical evolution of
the time-dependent GPEs. Initially the densities of the two components of
the mixture have the same profile, yielding a vanishing value of
magnetization. Once the domain wall moves, a nonvanishing magnetization is
formed, giving rise to a soliton which also exhibits a $2\pi$ jump in the
relative phase ($2\pi$ soliton). Thus the velocity plays the role of an
effective magnetic field, polarizing the soliton. As time evolves the
position of the soliton moves towards the periphery of the trapped gas and
increases its velocity as a consequence of the fact that its effective mass
is positive. Before reaching the border of the condensate, however, the soliton
slows down as a consequence of the fact that its effective mass at some
intermediate point, labeled ``B" in Fig.~\ref{fig:trapevolution}, changes
sign and becomes negative. Eventually the
soliton reaches zero velocity (labeled ``C" in Fig.~\ref%
{fig:trapevolution}) and is thereafter reflected towards the center of the
trap. When the Rabi coupling is much smaller than the critical value in Eq.~(%
\ref{eq:condition2}), soon after the inversion of the velocity, the $2\pi$
soliton exhibits a deep transformation characterized by a drastic change of
its phase and is transformed into a $0\pi$ soliton which does not exhibit an
asymptotic phase jump in the relative phase. This transformation takes place
when the local magnetization at the center of the soliton is equal to $1$,
which means that the density of one of the spin components exactly vanishes
(the X labeled ``D" in Fig.~\ref{fig:trapevolution}%
). The $0\pi$ soliton is then accelerated toward the center of the trap and
decelerated when it begins to reach the region of lower density, on the
opposite side of the trap, as a consequence of the negativity of its
effective mass. The $0\pi $ soliton cannot reach zero velocity and at some
point is transformed again into a $2\pi$ soliton which eventually reaches
zero velocity, to be reflected again. This highly non trivial dynamical
behavior is illustrated in Fig.~\ref{fig:trapevolution}, where the position
of the soliton is shown as a function of time.

The above concise description of the dynamics of magnetic solitons permits us
to understand the structure of the paper, which is organized as follows:
In Sec.~\ref{sec:formulation}, we formulate a variational approach to the
time-dependent GPEs, allowing for the identification of the solitonic
solutions. In Sec.~\ref{sec:staticsolution}, we derive analytic results for
the static and moving Son-Stephanov domain wall ($2\pi$ soliton) in the
presence of weak Rabi coupling. The general solutions of the $2\pi$ and $%
0\pi $ moving magnetic solitons are discussed in Sec.~\ref%
{sec:movingsolution}. The phase diagram and the properties of the magnetic
solitons are discussed in Sec.~\ref{sec:energyanddiagram}. Then we discuss
the dynamics and stability of the solitons in a one-dimensional (1D)
harmonic trap (Sec.~\ref{sec:1dtrap}) as well as in the presence of an
additional transverse confinement (Sec.~\ref{sec:2dtrap}). Section~\ref%
{sec:discuss} is devoted to the final discussion and conclusions.

\section{Solitons in uniform matter}

\label{sec:formulation}

\subsection{Equations for the magnetic solitons}

A two-component Bose-Einstein condensate in the presence of Rabi-coupling is
governed by two coupled GPEs which can be derived from the
Lagrangian density 
\begin{equation}
\mathcal{L}=\sum_{j=1}^2 \frac{i\hbar}{2} \left( \psi_j^*\frac{\partial}{%
\partial t}\psi_j - \psi_j\frac{\partial}{\partial t}\psi_j^*\right) - 
\mathcal{H},  \label{eq:lagrangiandensity}
\end{equation}
where $\psi_{j=1,2}$ are the wave functions of the two components and $%
\mathcal{H}$ is the Hamiltonian density given by 
\begin{eqnarray}
\mathcal{H} &=& \frac{\hbar^2}{2m}|\nabla\psi_1|^2+\frac{\hbar^2}{2m}%
|\nabla\psi_2|^2 -\frac{1}{2}\hbar\Omega(\psi_1^*\psi_2+\psi_2^*\psi_1) 
\notag \\
&& +\frac{g}{2}|\psi_1|^4+\frac{g}{2}|\psi_2|^4+g_{12}|\psi_1|^2|\psi_2|^2.
\label{eq:energydensity}
\end{eqnarray}
In this work, we have assumed the Rabi-coupling parameter $\Omega>0$. Under
conditions (\ref{eq:conditionOmega}) and (\ref{eq:condition1}),
the total density $n=n_1+n_2 $ of
the condensate exhibiting magnetic solitonic features can be assumed to
be constant~\cite{Qu2016}. As a consequence we can make the following ansatz
for the spinor order parameter,
\begin{equation}
\left( 
\begin{array}{c}
\psi_1 \\ 
\psi_2%
\end{array}
\right) = \sqrt{n} \left( 
\begin{array}{c}
\cos(\theta/2) e^{i\varphi_1} \\ 
\sin(\theta/2) e^{i\varphi_2}%
\end{array}
\right),  \label{eq:wavefunction}
\end{equation}
where $\varphi_{j=1,2}$ are the phases of the two wave functions. The
densities of the two components are given by $n_{1,2}=n(1\pm\cos\theta)/2$,
and the magnetization $m$ is calculated as $m=(n_1-n_2)/n= \cos\theta$.
Substituting Eq.~(\ref{eq:wavefunction}) into the Lagrangian density, Eq.~(%
\ref{eq:lagrangiandensity}), we obtain~\cite{Son2002} 
\begin{eqnarray}
\mathcal{L}&=&-n\hbar \left[ \cos^2\frac{\theta}{2}\frac{\partial\varphi_1}{%
\partial t} +\sin^2\frac{\theta}{2}\frac{\partial\varphi_2}{\partial t} %
\right]-\frac{n\hbar^2}{2m}\bigg[ \frac{1}{4}(\nabla\theta)^2  \notag \\
&& + \cos^2\frac{\theta}{2}(\nabla\varphi_1)^2+\sin^2\frac{\theta}{2}%
(\nabla\varphi_2)^2 \bigg]-\frac{1}{2}n^2 g  \notag \\
&& +\frac{1}{4}n^2\delta g\sin^2\theta +\frac{1}{2}n\hbar\Omega\sin\theta%
\cos(\varphi_1-\varphi_2).  \label{eq:Lag}
\end{eqnarray}
It is convenient to introduce the relative and total phases of the two
components 
\begin{eqnarray}
\varphi_A=\varphi_1-\varphi_2, \quad \varphi_B=\varphi_1+\varphi_2,
\label{eq:phases}
\end{eqnarray}
in terms of which, the Lagrangian density can be rewritten as 
\begin{eqnarray}
\mathcal{L}&=&-\frac{n\hbar }{2}\left( \cos \theta \partial _{t}\varphi
_{A}+\partial _{t}\varphi _{B}\right) -\frac{n\hbar ^{2}}{8m}\bigg[2\cos
\theta \nabla\varphi_A \nabla\varphi_B  \notag \\
&& +\left( \nabla\varphi _{A}\right) ^{2}+\left( \nabla\varphi _{B}\right)
^{2}+\left( \nabla\theta \right) ^{2}\bigg] -\frac{1}{2}n^2 g  \notag \\
&& +\frac{1}{4}n^2\delta g\sin^2\theta + \frac{1}{2}n\hbar\Omega\sin\theta%
\cos\varphi_A.  \label{Eq:lagrangiandensity2}
\end{eqnarray}%
It is important to note that the term $\partial_t\varphi_B$, as a
derivative, does not contribute to equations of motion and thus is
omitted in the following.

We begin our discussion by considering the 1D problem where all the
quantities depend only on the spatial coordinate $z$. We look for traveling
solutions of the form $\varphi _{A,B}=\varphi_{A,B}(z-Vt)$ and $\theta
=\theta (z-Vt)$ so that the Lagrangian density can be rewritten as 
\begin{eqnarray}
\mathcal{L} &=&\frac{n\hbar V}{2}\cos \theta \frac{\partial \varphi _{A}}{%
\partial z}-\frac{n\hbar ^{2}}{8m}\bigg[2\cos \theta \frac{\partial \varphi
_{A}}{\partial z}\frac{\partial \varphi _{B}}{\partial z}+\left( \frac{%
\partial \varphi _{A}}{\partial z}\right) ^{2}  \notag \\
&+&\left( \frac{\partial \varphi _{B}}{\partial z}\right) ^{2}+\left( \frac{%
\partial \theta }{\partial z}\right) ^{2}\bigg]-\frac{1}{2}n^2 g +\frac{1}{4}%
n^{2}\delta g\sin ^{2}\theta  \notag \\
&+&\frac{1}{2}n\hbar \Omega \sin \theta \cos \varphi _{A}.
\label{eq:lagrangiandensity3}
\end{eqnarray}%
It is instructive to reduce the Lagrangian density to a dimensionless form.
To this purpose, due to the magnetic nature of the solitons, the natural
units for the coordinates and velocities are chosen, respectively, as
the spin healing length and the spin sound velocity defined in the absence
of Rabi coupling: 
\begin{equation*}
\xi _{\text{s}}=\frac{\hbar }{\sqrt{2mn\delta g}},\qquad c_{\text{s}}=\sqrt{%
\frac{n\delta g}{2m}}.
\end{equation*}%
With the help of the following dimensionless variables for the position,
velocity and Rabi-coupling 
\begin{equation*}
\zeta =(z-Vt)/\xi _{\text{s}},\qquad U=V/c_{\text{s}},\qquad \omega _{\text{R%
}}=\frac{\Omega }{\Omega _{\text{c}}},
\end{equation*}%
the dimensionless Lagrangian density $\tilde{\mathcal{L}}=\mathcal{L}/nmc_{%
\text{s}}^{2}$ is given by 
\begin{eqnarray}
\tilde{\mathcal{L}} &=&U\cos \theta \frac{\partial \varphi _{A}}{\partial
\zeta }-\frac{1}{2}\bigg[\left( \frac{\partial \varphi _{A}}{\partial \zeta }%
\right) ^{2}+\left( \frac{\partial \varphi _{B}}{\partial \zeta }\right) ^{2}
\notag \\
&&+\left( \frac{\partial \theta }{\partial \zeta }\right) ^{2} +2\cos \theta 
\frac{\partial \varphi _{A}}{\partial \zeta }\frac{\partial \varphi _{B}}{%
\partial \zeta }\bigg] -\frac{g}{\delta g}  \notag \\
&& +\frac{1}{2}\sin ^{2}\theta +\frac{\omega _{\text{R}}}{3}\sin \theta \cos
\varphi _{A}.  \label{eq:lagrangiandensity4}
\end{eqnarray}%
Variation of the Lagrangian density with respect to the total phase $\varphi
_{B}$ gives 
\begin{equation}
\partial _{\zeta }\left( \frac{\partial \tilde{\mathcal{L}}}{\partial \left(
\partial _{\zeta }\varphi _{B}\right) }\right) =\partial _{\zeta }(\partial
_{\zeta }\varphi _{B}+\cos \theta \partial _{\zeta }\varphi _{A})=0.
\label{varfB}
\end{equation}%
We look for solitonic solutions by imposing the following boundary conditions: at $\zeta \rightarrow
\pm \infty $ the total and relative phases are constant and the spin is balanced (i.e., $\cos \theta =0$). Using the first boundary condition, we obtain the
equation
\begin{equation}
\frac{\partial \varphi _{B}}{\partial \zeta }+\cos \theta \frac{\partial
\varphi _{A}}{\partial \zeta }=0,  \label{dfBdz}
\end{equation}%
which, after substituting into Eq.~(\ref{eq:lagrangiandensity4}), yields 
\begin{eqnarray}
\tilde{\mathcal{L}} &=&U\cos \theta \frac{\partial \varphi _{A}}{\partial
\zeta }-\frac{1}{2}\left[ \left( \frac{\partial \theta }{\partial \zeta }%
\right) ^{2}+\sin ^{2}\theta \left( \frac{\partial \varphi _{A}}{\partial
\zeta }\right) ^{2}\right]  \notag \\
&&-\frac{g}{\delta g} +\frac{1}{2}\sin ^{2}\theta +\frac{1}{3}\omega _{\text{%
R}}\sin \theta \cos \varphi _{A}.  \label{eq:lagrangiansimple}
\end{eqnarray}%
The variation of $\tilde{\mathcal{L}}$ with respect to $\varphi _{A}$ and $%
\theta $ gives the two coupled differential equations for $\varphi_A$ and $%
\theta$~\cite{footnote}:
\begin{eqnarray}
&&\sin\theta\bigg( U \frac{\partial \theta }{\partial \zeta }+2 \cos \theta 
\frac{\partial \theta }{\partial \zeta }\frac{\partial \varphi _{A}}{%
\partial \zeta }+\sin \theta \frac{\partial ^{2}\varphi _{A}}{\partial \zeta
^{2}}  \notag \\
&& \qquad \qquad \qquad \qquad \qquad \qquad \qquad 
-\frac{\omega _{\text{R}}}{3}
\sin \varphi _{A} \bigg)=0,  \label{eq:ODE-phiA}  \\
&&-U\sin \theta \frac{\partial \varphi _{A}}{\partial \zeta }+\frac{\partial
^{2}\theta }{\partial \zeta ^{2}}-\sin \theta \cos \theta \left( \frac{%
\partial \varphi _{A}}{\partial \zeta }\right) ^{2}+\sin \theta \cos \theta 
\notag \\
&&\qquad \qquad \qquad \qquad \qquad +\frac{\omega _{\text{R}}}{3}\cos
\theta \cos \varphi _{A}=0.  \label{eq:ODE-theta}
\end{eqnarray}%
We point out that the same differential equations [i.e., Eqs.~(\ref{dfBdz}), (%
\ref{eq:ODE-phiA}), (\ref{eq:ODE-theta})] can also be derived by separating
the coupled GPEs into the real and imaginary parts. Furthermore,
Eqs.~(\ref{eq:ODE-phiA}) and (\ref{eq:ODE-theta}) are invariant under the
transformation
\begin{equation}
U\to -U, \quad \theta \to \pi-\theta,
\label{eq:transformation}
\end{equation}
i.e., $\cos\theta \to -\cos\theta$, $\sin\theta \to \sin\theta$. This transformation
allows us to obtain solutions for $U>0$ from solutions for $U<0$, and vice versa.
 Multiplying Eq.~(\ref%
{eq:ODE-phiA}) by $\partial \varphi _{A}/\partial \zeta $ and Eq.~(\ref%
{eq:ODE-theta}) by $\partial \theta /\zeta $ and then adding them together,
one can prove that the quantity 
\begin{eqnarray}
\tilde{\mathcal{G}} &=&-\frac{1}{2}\left[ \left( \frac{\partial \theta }{%
\partial \zeta }\right) ^{2}+\sin ^{2}\theta \left( \frac{\partial \varphi
_{A}}{\partial \zeta }\right) ^{2}\right]  \notag \\
&&+\frac{g}{\delta g} -\frac{1}{2}\sin ^{2}\theta -\frac{1}{3}\omega _{\text{%
R}}\sin \theta \cos \varphi _{A},  \label{eq:ODE-G}
\end{eqnarray}%
is position independent, i.e., $d\tilde{\mathcal{G}}/d\zeta =0.$ The boundary
conditions at $\zeta=\pm \infty$ imply $\tilde{\mathcal{G}}=(g/\delta
g-1/2-\omega_\text{R}/3)$. Taking this expression into account, we can
rewrite Eq.~(\ref{eq:ODE-G}) as 
\begin{eqnarray}
&&-\frac{1}{2}\left[ \left( \frac{\partial \theta }{\partial \zeta }\right)
^{2}+\sin ^{2}\theta \left( \frac{\partial \varphi _{A}}{\partial \zeta }%
\right) ^{2}\right]  \notag \\
&& +\frac{1}{2}\cos ^{2}\theta +\frac{1}{3}\omega _{\text{R}}(1-\sin \theta
\cos \varphi _{A})=0.  \label{eq:ODE-G2}
\end{eqnarray}

One can understand the physical origin of the integral $\tilde{\mathcal{G}}$
by noting that, if we consider $\zeta$ as a time variable, the quantity $%
\tilde{\mathcal{L}}$ in Eq.~(\ref{eq:lagrangiansimple}) is the \textit{%
time-independent} Lagrangian of a mechanical system with two degrees of
freedom, $\varphi_A$ and $\theta$. Then it is immediately clear that $\tilde{%
\mathcal{G}}$ is the \textit{conserving} energy of this auxiliary mechanical
system. It is important to stress, that $\tilde{\mathcal{G}}$ is different
from the actual energy density of the gas $\tilde{\mathcal{H}}=\mathcal{H}%
/nmc_\text{s}^2$ which can be obtained from $\tilde{\mathcal{G}}$ by
changing the sign of the first term in Eq.~(\ref{eq:ODE-G}).

As explored below, Eq.~(\ref{eq:ODE-G2}) provides a very useful
relation between the boundary conditions at infinity and those at $\zeta =0$. This
relation is crucial for the determination of the solutions
corresponding to the magnetic solitons.

\subsection{Energy of the magnetic solitons}
\label{sec:energy4mag}

For a moving magnetic soliton the analytical expression of its energy is not
accessible. However, the numerical solutions of the above differential
equations allow us to obtain the energy-velocity curve accurately, which is
crucial to understanding the physical properties of these solitons. As usual,
the energy of a magnetic soliton can be evaluated as the difference in
the canonical energies in the presence versus the absence of the soliton
(see Ref.~\cite{Pitaevskii2016}, Chap. 5). Thus we have 
\begin{eqnarray}
E &=&\frac{n\hbar c_{\text{s}}}{2}\int d\zeta \bigg[\bigg(\frac{1}{2}\left(
\partial \theta /\partial \zeta \right) ^{2}+\frac{1}{2}\sin ^{2}\theta
\left( \partial \varphi _{A}/\partial \zeta \right) ^{2}  \notag \\
&&+\frac{1}{2}\cos ^{2}\theta +\frac{1}{3}\omega _{\text{R}}(1-\sin \theta
\cos \varphi _{A})\bigg].  \label{E1}
\end{eqnarray}%
The integrand in the above equation is the difference in the dimensionless
energy densities $\tilde{\mathcal{H}}$, in the presence versus the absence
of the soliton. The derivative terms in Eq.~(\ref{E1}) can be eliminated
using Eq.~(\ref{eq:ODE-G2}). Finally, we find 
\begin{equation}
E =\frac{n\hbar c_\text{s}}{2}\int d\zeta \bigg[\cos^2\theta + \frac{2}{3}%
\omega_\text{R}(1-\sin\theta\cos\varphi_A)\bigg].  \label{E1simple}
\end{equation}
Once we find the solutions of the magnetic solitons, i.e., $\theta$ and $%
\varphi _{A}$, the corresponding soliton energy can be readily obtained by
integration. Although the velocity does not explicitly enter the above
equation, the energy of the soliton still depends on it since $\varphi_A$
and $\theta$ are velocity dependent. The effective mass, fixed by the
velocity dependence of the energy according to the definition 
\begin{equation}
m^{\ast }=\frac{1}{V}\frac{dE}{dV} \; ,  \label{m*}
\end{equation}%
can be extracted from the accurate numerical plot of the $E$-$V$ curve (see
Fig.~(\ref{fig:energy})).

\section{Analytical Results}

\label{sec:staticsolution} Analytic expressions for the magnetic solitons
can be obtained in special cases discussed in this section.

\subsection{Static Son-Stephanov domain wall}
\label{sec:SSsolution}
As the first example we recover the static Son-Stephanov domain wall solution
characterized by a relative phase jump of $2\pi$ in a spin-balanced system~%
\cite{Son2002}. By taking $U=0$ and $\theta=\pi/2$, the differential
equation, (\ref{eq:ODE-phiA}), for the relative phase becomes 
\begin{equation}
\frac{\partial^2\varphi_A}{\partial\zeta^2}-\frac{\omega_\text{R}}{3}%
\sin\varphi_A=0,  \label{eq:SSequation}
\end{equation}
which is the well-known sine-Gordon equation, whose solution is given by 
\begin{equation}
\varphi_A=4\arctan e^{\zeta \sqrt{\omega_\text{R}/3}}=4\arctan
e^{\kappa z},  \label{eq:SSsolution}
\end{equation}
with $\kappa=\sqrt{2m\Omega/\hbar}$ being the inverse of the characteristic
width of the relative phase domain wall,
\begin{equation}
\xi_\text{phase}=\kappa^{-1}=\xi_\text{s}\sqrt{\frac{3}{\omega_\text{R}}}=
\sqrt{\frac{\hbar}{2m\Omega}}.
\label{eq:Wallwidth}
\end{equation}

The analytic expression for the relative phase of the static domain wall
[see Eq.~(\ref{eq:SSsolution})] allows us to calculate the energy of the
solution explicitly. One finds: 
\begin{equation}
E_{\text{SS}}=4n\hbar c_\text{s}\sqrt{%
\frac{\omega_\text{R}}{3}} = 4n\hbar\sqrt{\frac{\hbar\Omega}{2m}},  \label{eq:energySS}
\end{equation}
and thus the dimensionless energy is $2E_\text{SS}/n\hbar c_\text{s}= 8\sqrt{%
\omega_\text{R}/3}$.

Son and Stephanov have proven that this solution corresponds to a local
minimum of the energy functional if condition (\ref{eq:condition2}) is
satisfied~\cite{Son2002}. In terms of dimensionless quantities,
condition (\ref{eq:condition2}) can be expressed as
$\omega_\text{R}<\omega_\text{R}^\text{c}\equiv 1$.
Note that there are two solutions for the static domain wall: one
exhibiting a $+2\pi$ phase jump and the other exhibiting a $-2\pi$ phase
jump. Moving magnetic domain walls can be developed from either of these
static domain walls and we focus on the solutions connected to the
former one.

Using the expression for the energy $E_\text{SS}$, one can justify our main assumption that
the total density is weakly affected by the presence of a magnetic
soliton. Let us consider a static domain wall. The number of
depleted atoms in the soliton can be calculated using the thermodynamic relation 
$N_\text{D}\equiv \int_{-\infty }^{\infty }\left[ n(z)-n\right] dz=-\partial{E}_\text{SS}%
/\partial \mu $, where $\mu =ng$ is the chemical potential. Straightforward calculation gives
$N_\text{D}=-4\hbar \sqrt{\hbar \Omega /(2mg^2)}$. One can thus estimate the
density perturbation near the center of the domain wall as $\left\vert
n(z)-n\right\vert \sim \left\vert N_\text{D}\right\vert /\xi _{\mathrm{phase}%
}\sim \hbar \Omega /g\ll n$ due to inequality (\ref{eq:conditionOmega})~\cite{footnote3}.

It is worthwhile discussing the connection of the above results to the well-known Manakov
limit solutions (i.e., when $\delta g=0$)~\cite{Manakov1974}. One can easily recognize that solution ~(\ref{eq:SSsolution}) is independent of the interaction strength $\delta g$. It is a formal solution of 
Eqs.~(\ref{eq:ODE-phiA}) and (\ref{eq:ODE-theta}) holding for any value of the
strength, including the value  $\delta g=0$. It is easy to check that in this case the simple rotation
\begin{equation}
\psi_{1,2}=(\psi_{D}\pm\psi_{B})/\sqrt{2}
\end{equation}
in spin space transforms Hamiltonian (\ref{eq:energydensity}) into the one used by Busch and Anglin~\cite{Busch2001}. Correspondingly, within the approximation $n=const$, solution (\ref{eq:SSsolution}) coincides with the static dark-bright soliton in Ref.~\cite{Busch2001}, when written in terms of the new variables. One should emphasize, however, that, as explicitly discussed in~\cite{Son2002}, this solution is energetically unstable. In the following we restrict ourselves mainly to configurations satisfying  the stability condition, (\ref{eq:condition2}), which requires $\delta g\neq 0$.

\subsection{Moving domain wall for weak Rabi coupling}
\label{Sec:slowDW}

The second example is a slowly moving domain wall ($2\pi$ magnetic soliton) whose
properties can be obtained analytically in the small Rabi-coupling limit,
\begin{equation}
\omega _{\text{R}}\ll 1.  \label{OR}
\end{equation}%
Under this condition, the width of the domain wall becomes much larger than
the spin healing length [see Eq.~(\ref{eq:Wallwidth})],
\begin{equation}
\xi _{\text{phase}}\gg \xi _{\text{s}},
\end{equation}%
and consequently, differentiation with respect to $\zeta $ gives a small
factor proportional to $\sqrt{\omega _{\text{R}}}$. Equation~(\ref{eq:ODE-theta})
then reduces to the simplified form: 
\begin{equation}
\cos \theta =U\partial _{\zeta }\varphi _{A}.  \label{costheta}
\end{equation}%
Integration with respect to $z$ gives a simple analytic expression for the
total magnetization 
\begin{equation}
\int_{-\infty}^{+\infty}\cos\theta dz = 2\pi \xi_s U.
\end{equation}
Substituting Eq.~(\ref{costheta}) into Eq.~(\ref{eq:ODE-phiA}), after
neglecting higher order terms, we obtain the differential equation for the
relative phase: 
\begin{equation}
\left( 1-U^{2}\right) \partial _{\zeta }^{2}\varphi _{A}-\frac{\omega _{%
\text{R}}}{3}\sin \varphi _{A}=0.
\end{equation}%
The similarity between this equation and the Son-Stephanov differential
equation [Eq.~(\ref{eq:SSequation})] indicates that all the results holding
at $U=0$ can be generalized to $U\neq 0$ by changing $\omega_{\text{R}}\to
\omega _{\text{R}}\left( 1-U^{2}\right) $ or $\Omega \to \Omega /\left(
1-U^{2}\right) $. In particular, the solution for the relative phase of the
moving domain wall is 
\begin{equation}
\varphi _{A}(U)=4\arctan \left[ \exp (\kappa(U)z )\right],  \label{son3}
\end{equation}%
with the width of the wall,
\begin{equation}
\xi_{\text{phase}}\left( U\right) =\kappa(U)^{-1}=\xi _{\text{s}}\sqrt{\frac{%
3(1-U^{2})}{\omega _{\text{R}}}},  \label{xiV}
\end{equation}%
becoming thinner and thinner as $U$ increases. With the help of Eq.~(\ref%
{costheta}) and Eq.~(\ref{son3}), one can calculate the energy [Eq.~\ref%
{E1simple}] of the moving domain wall. Ignoring higher order terms in $%
\omega_\text{R}$, one finds 
\begin{equation}
E(U)=4n\hbar\sqrt{\frac{\hbar\Omega}{2m(1-U^2)}}=4n\hbar c_s\sqrt{\frac{%
\omega_\text{R}}{3(1-U^2)}},
\label{eq:energySlowDW}
\end{equation}
which is actually the same expression for the energy of the static
Son-Stephanov domain wall [see Eq.~(\ref{eq:energySS})], with $\omega_\text{R}
$ replaced by $\omega_\text{R}/(1-U^2)$. Furthermore, using the definition
for the effective mass, we find 
\begin{equation}
m^{\ast}(U)=\frac{1}{V}\frac{dE}{dV}=\frac{4n\hbar }{c_{\text{s}}}\sqrt{%
\frac{\omega _{\text{R}}}{3}}\frac{1}{(1-U^2)^{3/2}}.
\end{equation}
Thus, the effective mass increases with an increase in $U$. However, we
emphasize that the equations derived in this section are not valid when $%
1-U^2$ is very small. For a low velocity one finds $m^{\ast}/m=8n\xi _{\text{%
s}}\sqrt{\omega _{\text{R}}/3}$. The positiveness of the effective mass
ensures the stability of the moving domain wall against snake instability. (For a general discussion of the snake instability for a plane soliton, see Ref.~\cite{Kamchatnov2008}).
It is noteworthy noticing that we derived the above analytical results under the
assumption in Eq.~(\ref{OR}). At a low velocity, the effective mass $m^{\ast }$ is, however, also positive
for finite values of $\omega _{\text{R}}$,
as long as $\omega _{\text{R}}< 1$. Actually, the fact that solution (\ref{eq:SSsolution}) corresponds to a local minimum
of the energy functional means that the energy increases for any small perturbations, including the one due a low velocity of motion. (See the numerical data in Fig.~\ref{fig:energy}.)

\section{Application of the theory:
General solutions for magnetic solitons}

\label{sec:movingsolution} As illustrated in Sec.~\ref{sec:intro}, after a
static Son-Stephanov domain wall is imprinted in a trapped binary
condensate, the domain wall starts moving and two types of solitons emerge
afterwards, oscillating in the trap. In this section, we obtain the exact
numerical solutions for both types of magnetic solitons in uniform matter.
Both solutions must satisfy the differential equations formulated in Sec.~%
\ref{sec:formulation}. However, different boundary conditions should be
imposed to identify the two different solutions. The difference in the
boundary conditions mainly affects the behavior of the relative phase.

\subsection{$2\protect\pi $ solitons}

The relative phase of these solitons exhibits the same $2\pi $ asymptotic
phase jump as in the static case. However, the spin population becomes
imbalanced in the wall center as soon as the velocity is different from
0. The boundary conditions for the $2\pi $ solitons are 
\begin{equation}
\theta (\zeta =\pm \infty )=\frac{\pi }{2},\varphi _{A}(\zeta =-\infty
)=0,\varphi _{A}(\zeta =+\infty )=2\pi ,  \label{eq:2pisolitonBC}
\end{equation}%
and we look for solutions characterized by the following symmetry
properties with respect to the wall center $\zeta =0$: 
\begin{equation}
\varphi _{A}(-\zeta )=2\pi -\varphi _{A}(\zeta ),\qquad \theta (-\zeta
)=\theta (\zeta ),  \label{eq:sym1}
\end{equation}%
which implies $\varphi _{A}(0)=\pi ,$ $\partial _{\zeta }\theta |_{\zeta
=0}=0$. With the help of Eq. (\ref{eq:ODE-G2}), a relation between the
boundary conditions for $\varphi _{A}$ and those for $\theta $ at $\zeta =0$ can be
established and hence one finds the slope of the relative phase as 
\begin{equation}
\left( \frac{\partial \varphi _{A}}{\partial \zeta }\right) _{\zeta =0}^{2}=%
\frac{\cos ^{2}\theta _{0}+\frac{2}{3}\omega _{\text{R}}(1+\sin \theta _{0})%
}{\sin ^{2}\theta _{0}}~,  \label{eq:slope}
\end{equation}%
where the value of $\theta _{0}=\theta (\zeta =0)$ determines the
magnetization at the center of the soliton: $m_{0}\equiv m(\zeta =0)=\cos
\theta _{0}$. Equation~(\ref{eq:slope}), a direct consequence of the
boundary conditions at $\zeta =\pm \infty $, is important because it
provides a boundary condition at $\zeta =0$ which is much more useful in
order to find the solitonic solutions rather than fixing the boundary
conditions at infinity.

The procedure to find the solutions of the coupled differential
equations, Eq.~(\ref{eq:ODE-phiA}) and Eq.~(\ref{eq:ODE-theta}), is the
following: For a given velocity $U$ and Rabi-coupling strength $\omega _{%
\text{R}}$, we carefully tune the input parameter $\theta _{0}$ until the
solutions of these differential equations converge to a form satisfying the
boundary conditions in Eq.~(\ref{eq:2pisolitonBC}) for the magnetic solitons~%
\cite{footnote2}. The two possible signs for the slope are related to the
two static Son-Stephanov domain wall solutions as $U\rightarrow 0$ and $%
m_0\to 0$.

\begin{figure}[tbp]
\includegraphics[width=8.0cm]{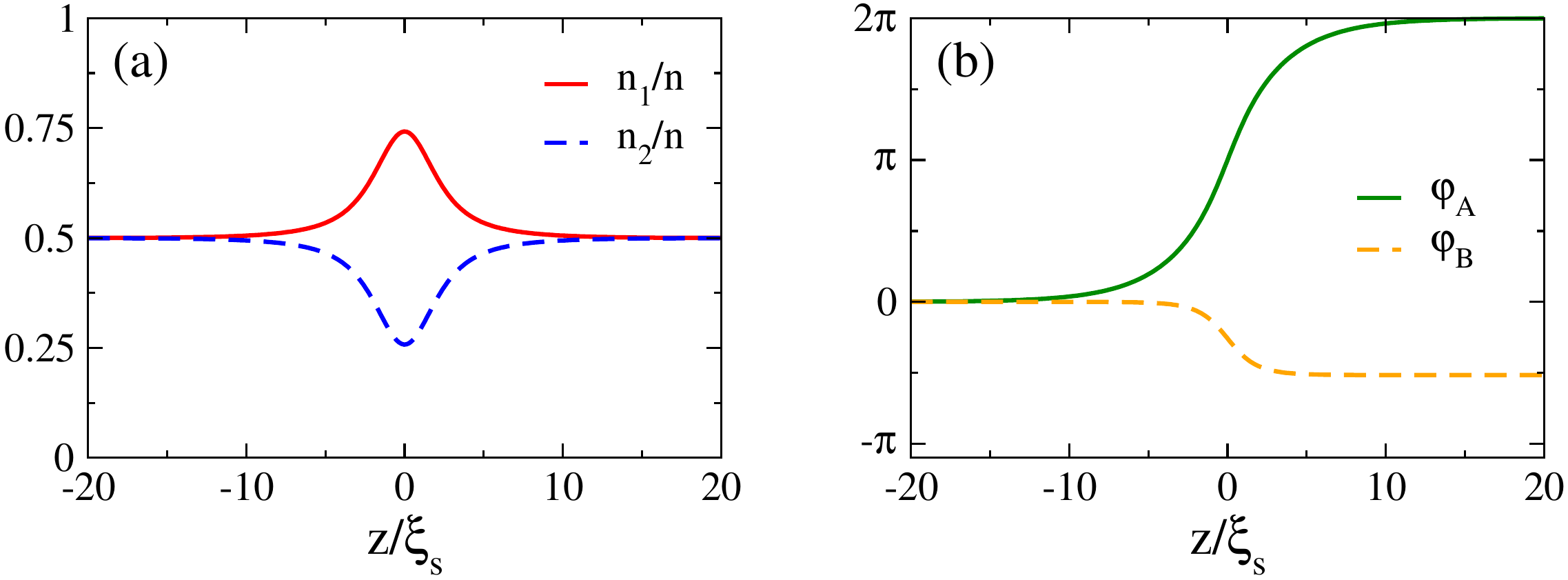}
\caption
{Profiles of the $2\protect\pi$ soliton with velocity $V/c_\text{s}%
=0.28$, $m_0=0.48$, and Rabi coupling $\protect\omega_\text{R}=0.3$. (a) Solid red and dashed blue
lines represent the density distributions of the two
components, satisfying $(n_1+n_2)/n=1$. (b) Solid green and dashed orange
lines show the relative phase $\protect\varphi_A$ and total phase $%
\protect\varphi_B$ as a function of the coordinate. The jump of the relative
phase is $2\protect\pi$, independent of the velocity. This solution is close
to the critical velocity where the effective mass diverges (see discussion
in Sec.~\protect\ref{sec:energyanddiagram}).}
\label{fig:typeI}
\end{figure}

\begin{figure}[tbp]
\includegraphics[width=8.0cm]{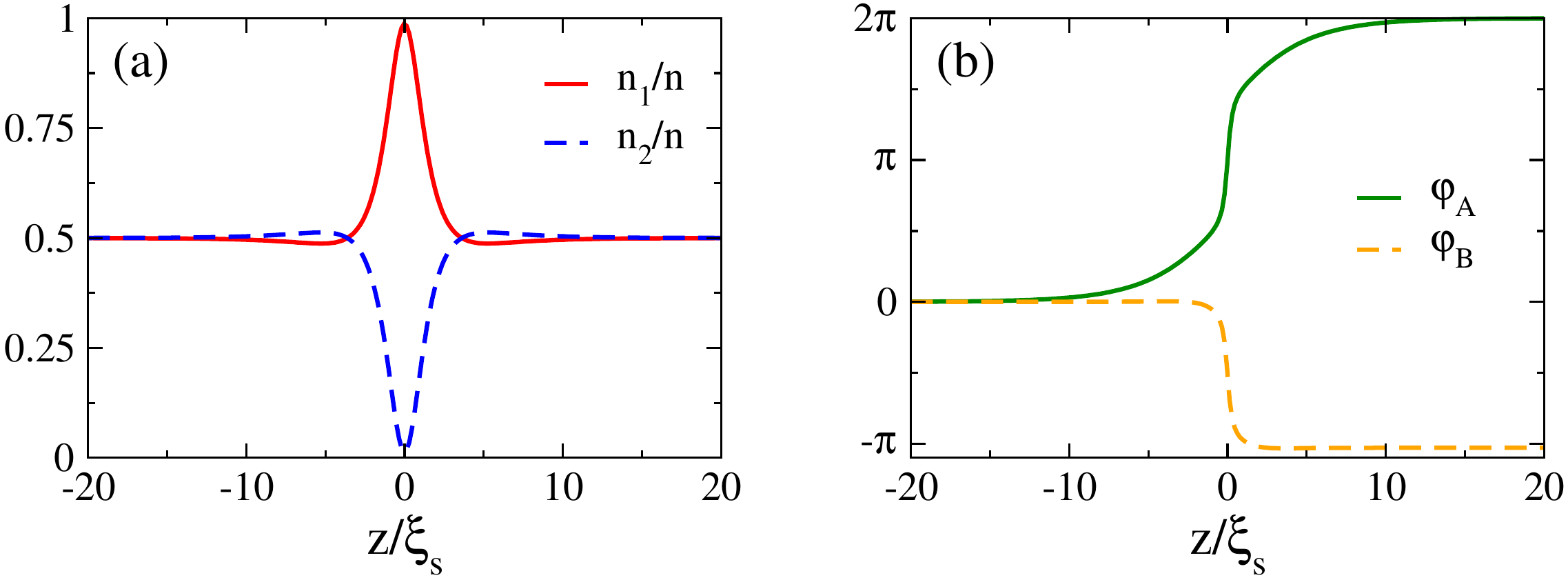}
\caption{Profiles of a $2\protect\pi$ soliton with velocity $V/c_\text{s}%
=-0.25$, $m_0=0.97$, and Rabi coupling $\protect\omega_\text{R}=0.3$. (a) Solid
red and dashed blue lines represent the density distributions of the two
components, satisfying $(n_1+n_2)/n=1$. Note that two nodes appear at the
wings of the soliton. (b) Solid green and dashed orange lines show the
relative phase $\protect\varphi_A$ and total phase $\protect\varphi_B $ as a
function of the coordinate. This soliton has a negative effective mass and
corresponds to the solution very close to the green X with $m_0=1$
in Fig.~\protect\ref{fig:energy}.}
\label{fig:typeIwithnodes}
\end{figure}

Figures~\ref{fig:typeI} and ~\ref{fig:typeIwithnodes} show the density
distributions and the relative and total phases of a $2\pi $ soliton with
positive ($U=V/c_{\text{s}}=0.28$) and negative ($U=V/c_{\text{s}}=-0.25$)
velocities, respectively. The difference between the two cases is that they
correspond, respectively, to a solution before and one after the turning point
(see Fig.~\ref{fig:trapevolution}). The latter case is characterized by a
much higher magnetization (close to $1$). For a negative velocity with even
larger $|U|$ (longer evolution times in Fig.~\ref{fig:trapevolution}), the
density of one component vanishes at $\zeta =0$ and the $2\pi $ soliton
breaks off, being converted into a $0\pi $ soliton.

\subsection{$0\protect\pi $ solitons}

Let us now discuss the main features of $0\pi $ solitons. Our results, based
on GPE simulations, show that a $2\pi $ soliton transforms into a $0\pi $
soliton when the density of one component vanishes at $\zeta =0$ where its
phase is not well defined and thus can change by $2\pi $ without any energy
cost. Although the asymptotic $2\pi $ phase jump disappears, the relative
phase still varies as a function of the position. The boundary conditions now
become 
\begin{equation}
\theta (\zeta =\pm \infty )=\frac{\pi }{2}~,\quad \varphi _{A}(\zeta =\pm
\infty )=0,  \label{eq:0pisolitonBC}
\end{equation}%
and the natural symmetries of the $\varphi $ and $\theta $ functions are 
\begin{equation}
\varphi _{A}(-\zeta )=-\varphi _{A}(\zeta )~,\quad \theta (-\zeta )=\theta
(\zeta ),  \label{eq:sym2}
\end{equation}%
which implies $\varphi _{A}(0)=0,\partial _{\zeta }\theta |_{\zeta =0}=0$.
Using Eq.~(\ref{eq:ODE-G2}), analogously to the derivation of Eq.~(\ref%
{eq:slope}), we obtain the slope of the relative phase at the soliton center
as 
\begin{equation}
\left( \frac{\partial \varphi _{A}}{\partial \zeta }\right) _{\zeta =0}^{2}=%
\frac{\cos ^{2}\theta _{0}+\frac{2}{3}\omega _{\text{R}}(1-\sin \theta _{0})%
}{\sin ^{2}\theta _{0}},  \label{magnSolrelPhaseSlope}
\end{equation}%
where $\theta _{0}=\theta (\zeta =0)$ determines the magnetization of the $%
0\pi $ soliton at $\zeta =0$.

The procedure for finding the solutions is similar to the one developed in
the previous section: for a given velocity $U$ and Rabi-coupling strength $%
\omega_\text{R} $, we can tune $\theta_0$ until the solution of the above
differential equations is consistent with the boundary conditions in Eq.~(%
\ref{eq:0pisolitonBC}).

\begin{figure}[tbp]
\includegraphics[width=8.0cm]{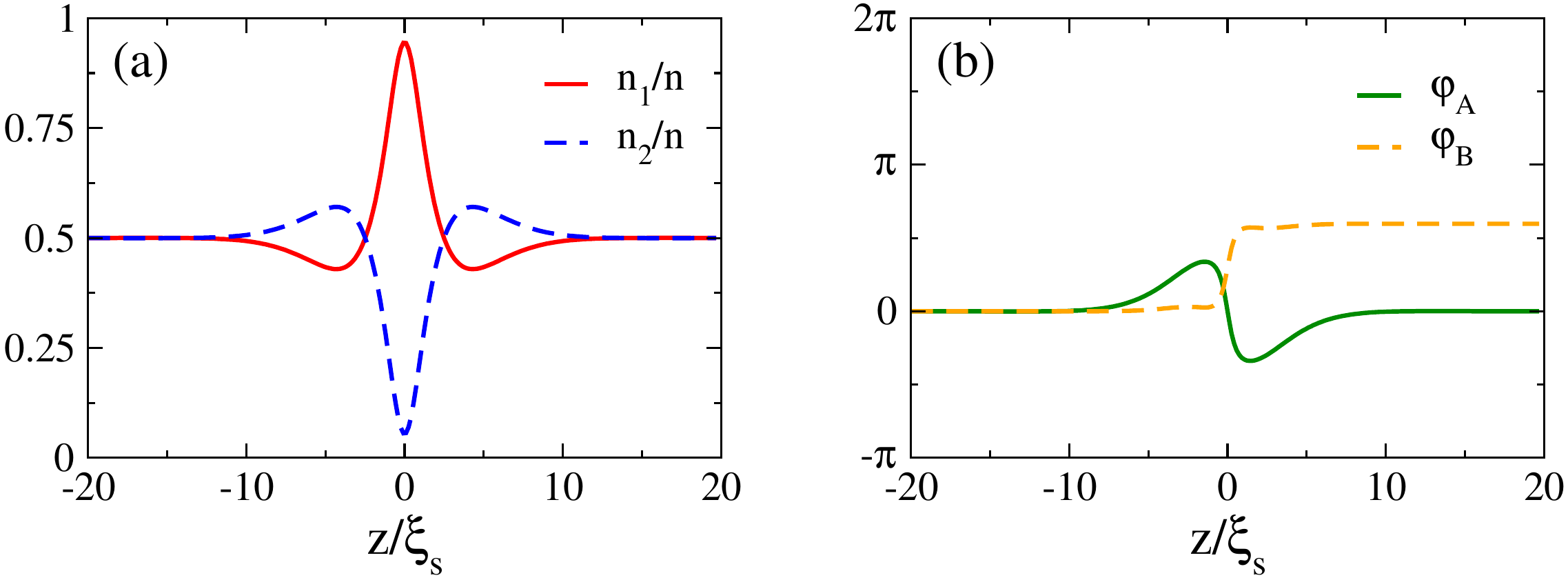}
\caption{Profiles of a $0\protect\pi$ soliton with velocity $V/c_\text{s}%
=-0.9$, $m_0=0.89$, and Rabi coupling $\protect\omega_\text{R}=0.3$. (a) Solid
red and dashed blue lines show the density distributions of the two
components, satisfying $(n_1+n_2)/n=1$. (b) Solid green and dashed orange
lines show the relative phase $\protect\varphi_A$ and total phase $%
\protect\varphi_B$ as a function of the coordinate. This $0\protect\pi$
soliton has a negative effective mass and the asymptotic jump of $\protect%
\varphi_A$ is 0.}
\label{fig:typeII}
\end{figure}

Figure~\ref{fig:typeII} shows the profile of a $0\pi$ soliton with negative
velocity $U=V/c_{\text{s}}=-0.9$. The density is magnetized in the solitonic
region and has two spin-balanced points followed by two oppositely
magnetized regions on the wings. We remind that the relative phase of the $%
0\pi $ soliton is an odd function of $\zeta $ and does not exhibit any
asymptotic phase jump. Below we show that, as the velocity increases,
more and more oscillations appear in the profile of $0\pi$ solitons.

\subsection{Transformation between $2\protect\pi$ and $0\protect\pi$ solitons}

\begin{figure}[tbp]
\includegraphics[width=7cm]{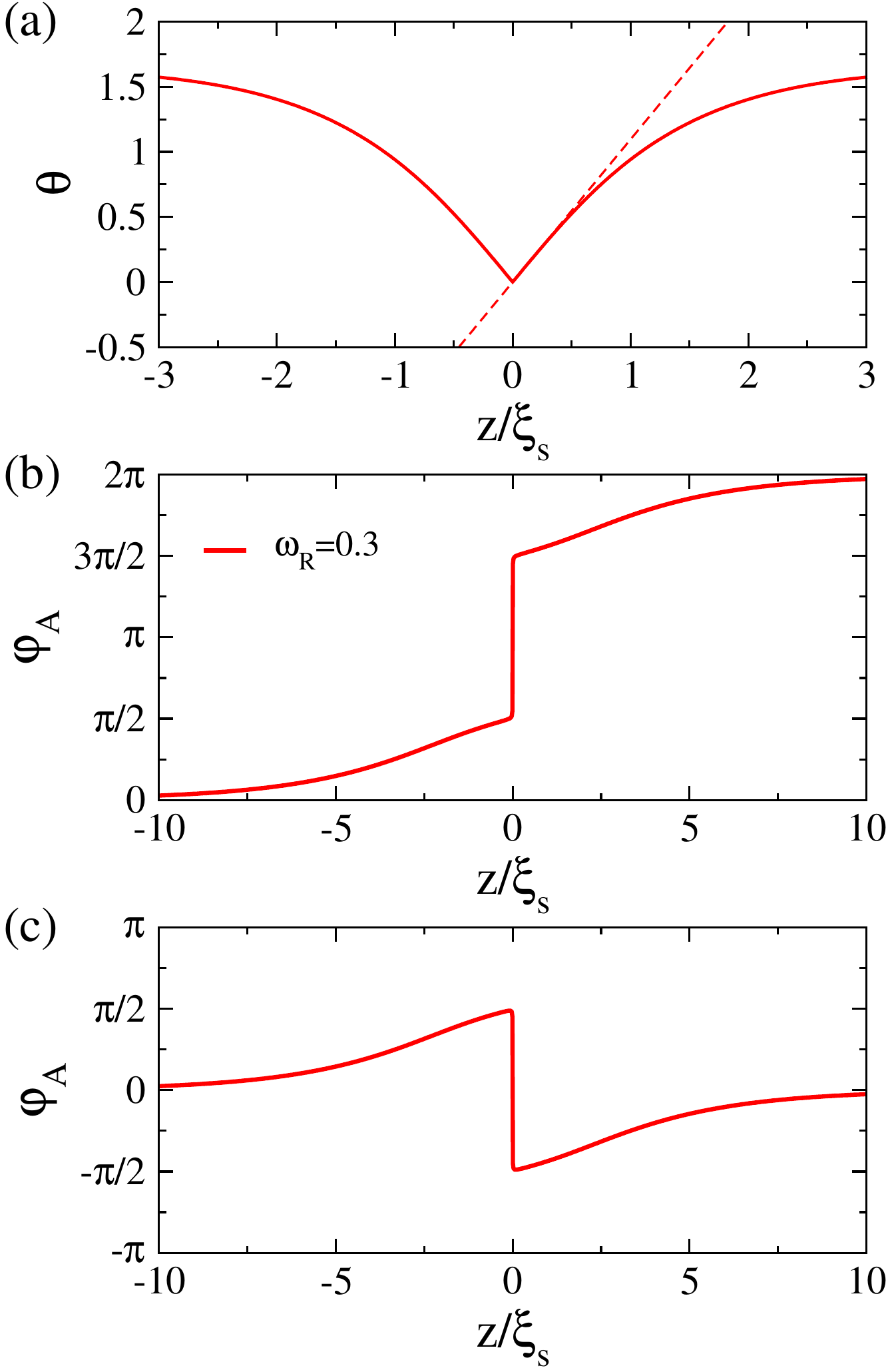}
\caption{(a) Plot of $\protect\theta (\protect\zeta )$ at the transformation
point between $2\protect\pi $ and $0\protect\pi $ solitons for
Rabi coupling strength $\protect\omega _{\text{R}}=0.3$. 
(a) Dashed lines show the analytical
prediction of the behavior of $\protect\theta (\protect\zeta )$ at small $%
\protect\zeta \rightarrow 0$ [see Eq.~(\protect\ref{eq:thetaslope})]. Relative
phases (b) before and (c) after the transformation, which exhibit a
$2\protect\pi $ and $0\protect\pi $ phase jump,
respectively. Velocities of the solitons are $U=-0.464$ (b) and $U=-0.467$ (c). 
As $\zeta\to -0$, the relative phase $\varphi_A \to \pi/2$ in (b,c).
As $\zeta \to +0$, $\varphi_A \to 3\pi/2$ (b) and $\varphi_A \to -\pi/2$ (c). }
\label{fig:transformation}
\end{figure}

In this subsection we discuss the solution at the transformation point
between $2\pi$   and $0\pi $ solitons. These two solitons
have different symmetry properties due to the $2\pi$ jump in the relative phase
$\varphi_A$ of $2\pi$ solitons, which cannot disappear continuously, but only
in a discrete way where the density of one
of the two components vanishes at $\zeta=0$. This is a singular point and
requires  a separate investigation. We consider the case where $n_{2}(\zeta =0)=0$
and thus $\theta _{0}=\theta (\zeta =0)=0$, i.e.,  $m_{0}=\cos \theta _{0}=1$. 
As shown in the following section, the transition between the two solitons with $n_2(\zeta=0)=0$ takes place at a negative velocity $U<0$. The transition for $U>0$ with $n_1(\zeta=0)=0$ can be obtained using the transformation
in Eq.~(\ref{eq:transformation}).
Our numerical calculations clearly show that the function $\theta(\zeta)$ vanishes at $\zeta=0$
with a finite slope (see Fig.~\ref{fig:transformation}(a)). The slope of $\theta(\zeta)$
at $\zeta=0$ can be analytically derived from Eq.~(\ref{eq:ODE-G2}) and one finds 
\begin{equation}
\left( \frac{\partial \theta }{\partial \zeta }\right)
 \bigg|_{\zeta \to \mp 0}=\mp \sqrt{1+\frac{2}{3}\omega _{R}},  \label{eq:thetaslope}
\end{equation}%
which is in good agreement with the numerical results [see  Fig.~\ref{fig:transformation}(a)].

Let us now discuss the behavior of $\varphi _{A}$ near the transformation point. 
We consider a solution, where $\varphi _{A}\rightarrow 0$ at $\zeta
\rightarrow -\infty $.\
The numerical calculation shows that in this case the relative phase $\varphi _{A}(\zeta
=0)$ approaches the value $\pi /2$ as $\zeta \to -0$.

According to Eq.~(\ref{eq:ODE-phiA}) we have, for all $\zeta \neq 0$,  \ 
\begin{equation}
\left( U\frac{\partial \theta }{\partial \zeta }+2\cos \theta \frac{\partial
\theta }{\partial \zeta }\frac{\partial \varphi _{A}}{\partial \zeta }+\sin
\theta \frac{\partial ^{2}\varphi _{A}}{\partial \zeta^2 }-\frac{\omega _{R}}{3%
}\sin \varphi _{A}\right) =0\ .
\end{equation}%
As $\zeta \to -0$ this equation gives  
\begin{equation}
\left( U\frac{\partial \theta }{\partial \zeta }+2\frac{\partial \theta }{%
\partial \zeta }\frac{\partial \varphi _{A}}{\partial \zeta }-\frac{\omega
_{R}}{3}\right) \bigg|_{\zeta \rightarrow -0}=0
\end{equation}%
or, taking Eq.~(\ref{eq:thetaslope}) into account, 
\begin{equation}
\left( \frac{\partial \varphi _{A}}{\partial \zeta }\right) \bigg|_{\zeta\to -0}=-\frac{1}{2}\left[ U+\frac{\omega _{R}}{\sqrt{9+6\omega _{R}}%
}\right] .  \label{eq:phiAslope}
\end{equation}%
Result (\ref{eq:phiAslope}) for the slope of the relative phase is confirmed by our calculations with a good accuracy. 

At the transformation point, Eqs. (\ref{eq:thetaslope}) and (\ref{eq:phiAslope})
replace Eqs. (\ref{eq:slope}) and (\ref{magnSolrelPhaseSlope}%
). The solutions at $\zeta \leq 0$ are the same for both $2\pi $ and $0\pi $
solitons. The solution at $\zeta >0$ can be obtained using
the symmetry properties (\ref{eq:sym1}) and (\ref%
{eq:sym2}) for the $2\pi $ and $0\pi $ solitons, respectively. Distributions of the angle $\theta $ are the
same for two types of solitons [see Fig. \ref{fig:transformation}
(a)]. Distributions of the relative phase for $2\pi$ and $0\pi$ solitons
 are shown in Figs.~\ref%
{fig:transformation} (b) and ~\ref{fig:transformation}(c), respectively.
The presented solutions obviously
satisfy the correct boundary conditions at $\zeta \rightarrow \infty $. 

The solutions at the transformation point exhibit a strong singularity at $%
\zeta =0$: a $\pi $-jump in the relative phase $\varphi _{A}$ and a jump in $%
\partial \theta /\partial \zeta $. However, this singularity can be
eliminated by a simple substitution of variables at $\zeta >0$. Let us
define $\theta (\zeta )=\nu (\zeta )$ at $\zeta <0$ and $\theta (\zeta
)=-\nu (\zeta )$ at $\zeta >0.$ Analogously, let at $\zeta <0,$ $\varphi
_{A}(\zeta )=\chi _{A}(\zeta )$ and at $\zeta >0, $let $\varphi _{A}(\zeta
)=\chi _{A}(\zeta )-\pi $ for $2\pi $ soliton and $\varphi _{A}(\zeta )=\chi
_{A}(\zeta )+$ $\pi $ for $0\pi $ solitons. The new functions $\nu
(\zeta )$ and $\chi _{A}(\zeta )$ have no singularities at $\zeta =0$. They
satisfy the symmetry conditions 
\begin{equation}
\nu (-\zeta )=-\nu (\zeta ),\quad \chi _{A}(-\zeta )=\pi -\chi _{A}(\zeta )\ .
\end{equation}%
Actually, the functions $\nu (\zeta )$ and $\chi _{A}(\zeta )$ are analytic
continuations of the functions $\theta (\zeta )$ and $\varphi _{A}(\zeta )$ from 
$\zeta <0$ to  all  values of $\zeta $. Before concluding this section,
we point out that phase reduction is not a unique property of our system. For example, phase
slip has been
observed by moving a weak link in a toroidal atomic BEC~\cite{Wright2013}.

\section{Phase diagram and properties of magnetic solitons}

\label{sec:energyanddiagram}

\subsection{Magnetization and energy}

\begin{figure}[tbp]
\includegraphics[width=7.5cm]{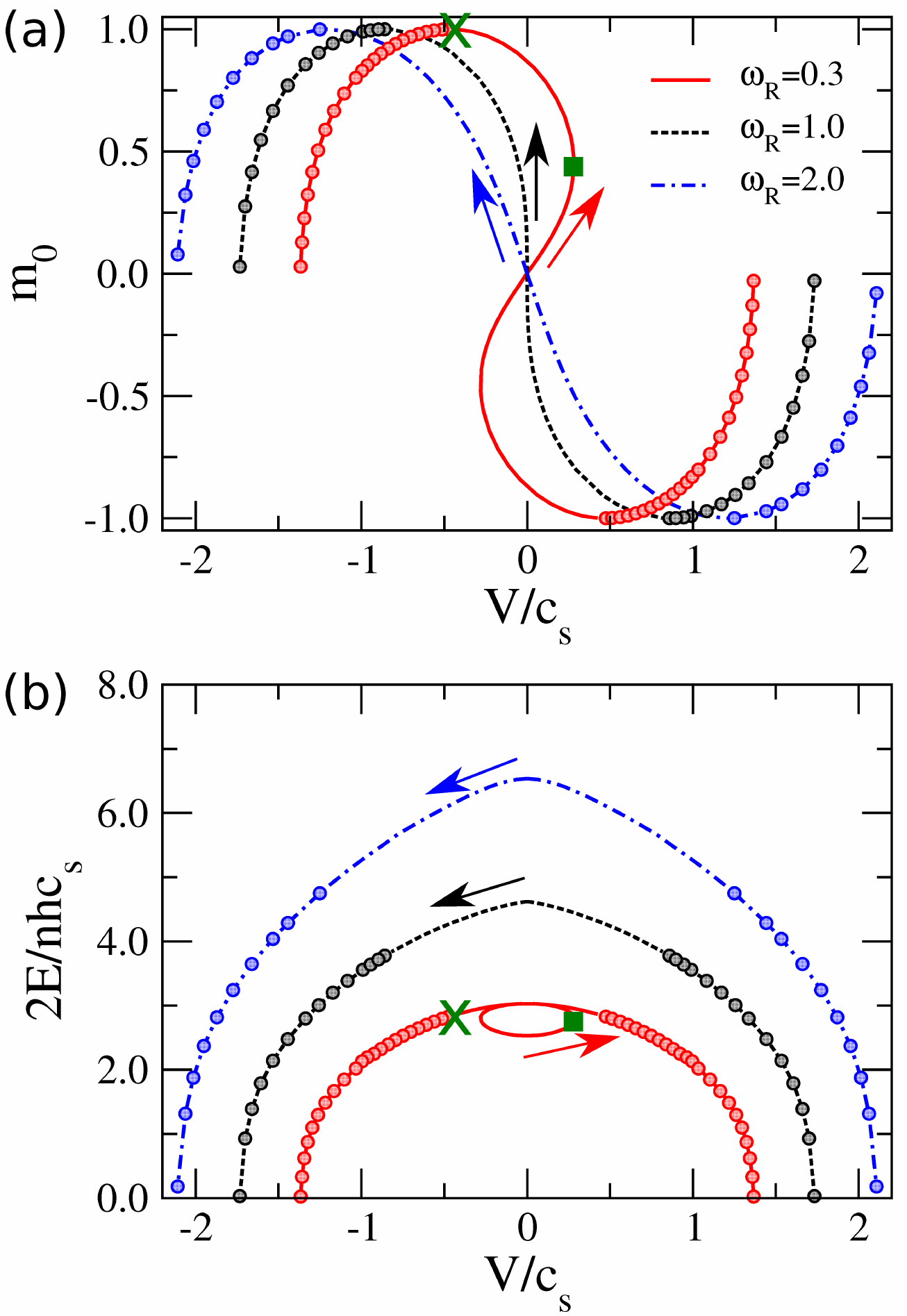}
\caption{(a) Phase diagram of magnetic solitons in the $m_0$-$U$ plane,
where $m_0$ is the magnetization at the center of the soliton and $U=V/c_%
\text{s}$ is the velocity. (b) Velocity dependence of the energy of magnetic
solitons for different Rabi coupling strengths $\protect\omega_\text{R}=0.3$%
(red solid line), $1$ (dashed black line) and $2$ (dash-dotted blue line).
Lines without circles indicate that the solutions
are the $2\protect\pi$ solitons; lines with circles, that
the solutions are $0\protect\pi$ solitons. The point of origin in (a) corresponds
to the solution of the known static Son-Stephanov domain
wall with a $+2\protect\pi$ relative phase jump and its energy increases as
the Rabi coupling increases [see (b)]. The green square
indicates the solution where the effective mass of the $2\protect\pi$
soliton diverges and the green X indicates the position of the
transformation between $2\protect\pi$ and $0\protect\pi$ solitons for $%
\protect\omega_\text{R}=0.3$. Note that there exists another series of
solutions obtained by changing (a) according to the
transformation $V\rightarrow -V$, then the solutions are connected to the
known static Son-Stephanov domain wall with a $-2\protect\pi$ relative phase
jump.}
\label{fig:energy}
\end{figure}

Our main results are presented in Fig.~\ref{fig:energy} where we show the
curves for three values of $\omega_\text{R}$: $\omega_\text{R}=0.3$%
, $1$, and $2$, which correspond to less than, equal to, and larger than the
critical value, (\ref{eq:condition2}), for the Rabi coupling below which the
Son-Stephanov solution for the domain wall is stable. In Figs.~\ref{fig:energy}(a) and~\ref{fig:energy}(b),
solid lines without circles label the results for $2\pi$ solitons which
exhibit a $2\pi$ relative phase jump, while solid lines with circles
label $0\pi$ solitons which do not exhibit an asymptotic relative phase jump.

It is easy to recognize that the origin of Fig.~\ref{fig:energy}(a), the
solution with $U=0$ and $m_0=0$, is the Son-Stephanov static domain wall. As
shown by the red curve in Fig.~\ref{fig:energy}(b), this solution is a local
minimum of the $E$-$V$ curve as long as $\omega_\text{R}< 1$. The effective
mass of solitons [see Eq.~(\ref{m*})] is related to the slope of the $E$-$V$
line. As shown in Fig.~\ref{fig:energy}(b), the effective mass of a $2\pi$
soliton can be positive or negative when the Rabi coupling is smaller than
the critical value ($\omega_\text{R}<1$), while it is always negative when $%
\omega_\text{R}\geq 1$. In contrast, the effective mass of $0\pi$ soliton is
always negative, irrespective of the strength of the Rabi coupling. Note that $%
2\pi$ solitons with a positive effective mass are not affected by snake
instability.

Let us now discuss in more detail the phase diagram in Fig.~\ref{fig:energy}%
.

(i) $\omega_\text{R}<1$. This is the most interesting case, where a $2\pi$
magnetic soliton with a positive effective mass is predicted to exist. Moving
continuously from the solution at the origin ($U=0$) in Fig.~\ref{fig:energy}%
(a), the solution exists also for finite values of $U$ and is associated
with a positive effective mass and a finite value of the magnetization (red
arrow and its opposite direction). The effective mass of such solutions
diverges at a critical value of the velocity (indicated by the green square
in the figure). The profiles for the densities and phases at this critical
point are shown in Fig.~\ref{fig:typeI}. $2\pi$ solitonic solutions
with larger values of $|U|$ do not exist. However, $2\pi$ solitons with
smaller $|U|$ and larger magnetization exist as clearly shown in Fig.~\ref%
{fig:energy}(a), their effective mass becoming negative. 
For even larger magnetization, $2\pi$ solitonic solutions with
opposite velocity exist where two nodes
appear on the wings of the soliton as shown in Fig.~\ref%
{fig:typeIwithnodes}. The nonmonotonic dependence of the magnetization on
the velocity of the soliton [see Fig.~\ref{fig:energy}(a)] is responsible
for the loop of the energy as a function of $V$ in the same interval of
velocities [see Fig.~\ref{fig:energy}(b)].

When $|m_{0}|=1$, i.e.. when the density of one component vanishes (green X in the figure), the
corresponding phase is not defined. Then the $2\pi $ relative phase jump
disappears and a $0\pi $ solitonic solution (solid line with circles emerges
at larger $|U|$). The profiles of the density and of the phases of typical $%
0\pi$ solitonic solutions are shown in Fig.~\ref{fig:typeII}. The $0\pi $
solitonic solution continues by increasing the velocity with the
corresponding decrease in magnetization until it reaches a critical velocity 
$U_\text{L}=V_\text{L}/c_{\text{s}}$, where the solitonic solution
disappears, its energy approaching 0 (see below).

\begin{figure}[tbp]
\includegraphics[width=6.0cm]{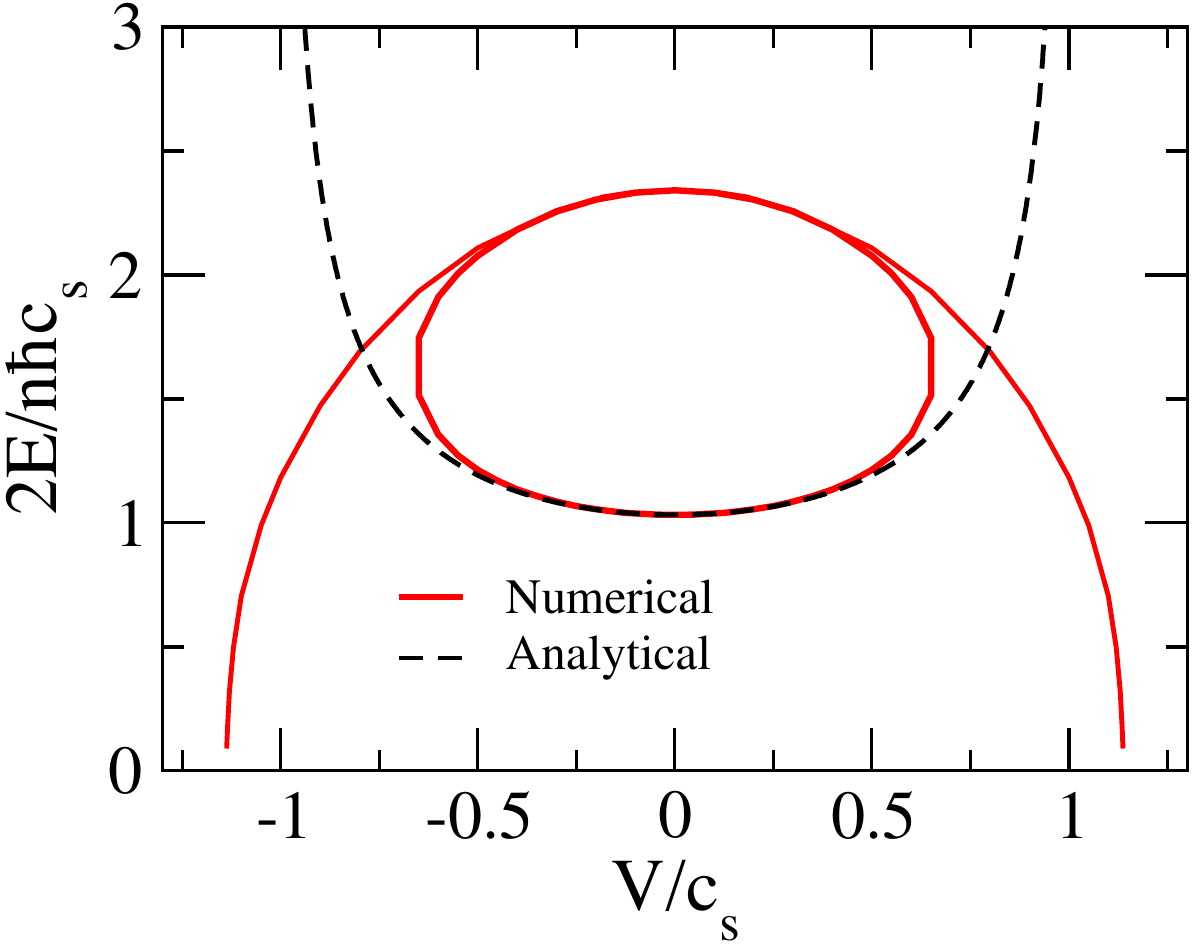}
\caption{Comparison of the numerically calculated energy and the theoretical prediction for a slowly moving domain wall in the presence of a weak Rabi coupling $\omega_\text{R}=0.05$. The dashed black line shows the analytic result [Eq.~(\ref{eq:energySlowDW})] and red lines show the numerical results.}
\label{fig:energySlowDW}
\end{figure}

We can now check the validity of our theory for a slowly moving domain wall ($2\pi$ magnetic soliton) developed in Sec.~\ref{Sec:slowDW}. In Fig.~\ref{fig:energySlowDW}, we compare the energy of the magnetic soliton numerically calculated in the presence of a small Rabi coupling $\omega_R=0.05$ and that from the analytical expression given by Eq.~(\ref{eq:energySlowDW}). As anticipated, we find that they agree with each other very well at low velocities.

(ii) $\omega_\text{R}\geq 1$. Despite the fact that the stability of moving solitons for $\omega_R\ge 1$ remains an unexplored problem, in
Fig.~\ref{fig:energy} we present the velocity dependence resulting from the formal solution of our coupled differential equations [Eqs.~(\ref{eq:ODE-phiA} and ~\ref{eq:ODE-theta})] for
$\omega_R=2$. The corresponding solitons are characterized by a monotonic behavior of the
magnetization as a function of the velocity [see blue curve in Fig.~\ref%
{fig:energy}(a)]. The energy of the soliton decreases when $|U|$ increases,
corresponding to a negative effective mass [see blue curves in Fig.~\ref%
{fig:energy}(b)]. One should however take into
account that these solitons, according to the findings of~\cite{Son2002}, are unstable for low enough velocities.

The case $\omega_\text{R}=1$ [see black curve in Fig.~\ref{fig:energy}(a)],
corresponding to the boundary of stability of the Son-Stephanov domain wall,
is a special one. At this value of $\omega_\text{R}$, the ``polarizability" $%
d(m_0)/dU\to \pm\infty$ when $U\to \pm 0$. The singularity on the black
curve at $U=0$ in Fig.~\ref{fig:energy}(b) is related to this divergence.

Further investigation of these solutions should concern their stability at finite $U$. The investigation
of this problem, however, lies beyond the scope of this work, which addresses mainly the $\omega_\text{R}<1$ case.

\subsection{Landau critical velocity of $0\protect\pi$ solitons}

\begin{figure}[tbp]
\includegraphics[width=6.0cm]{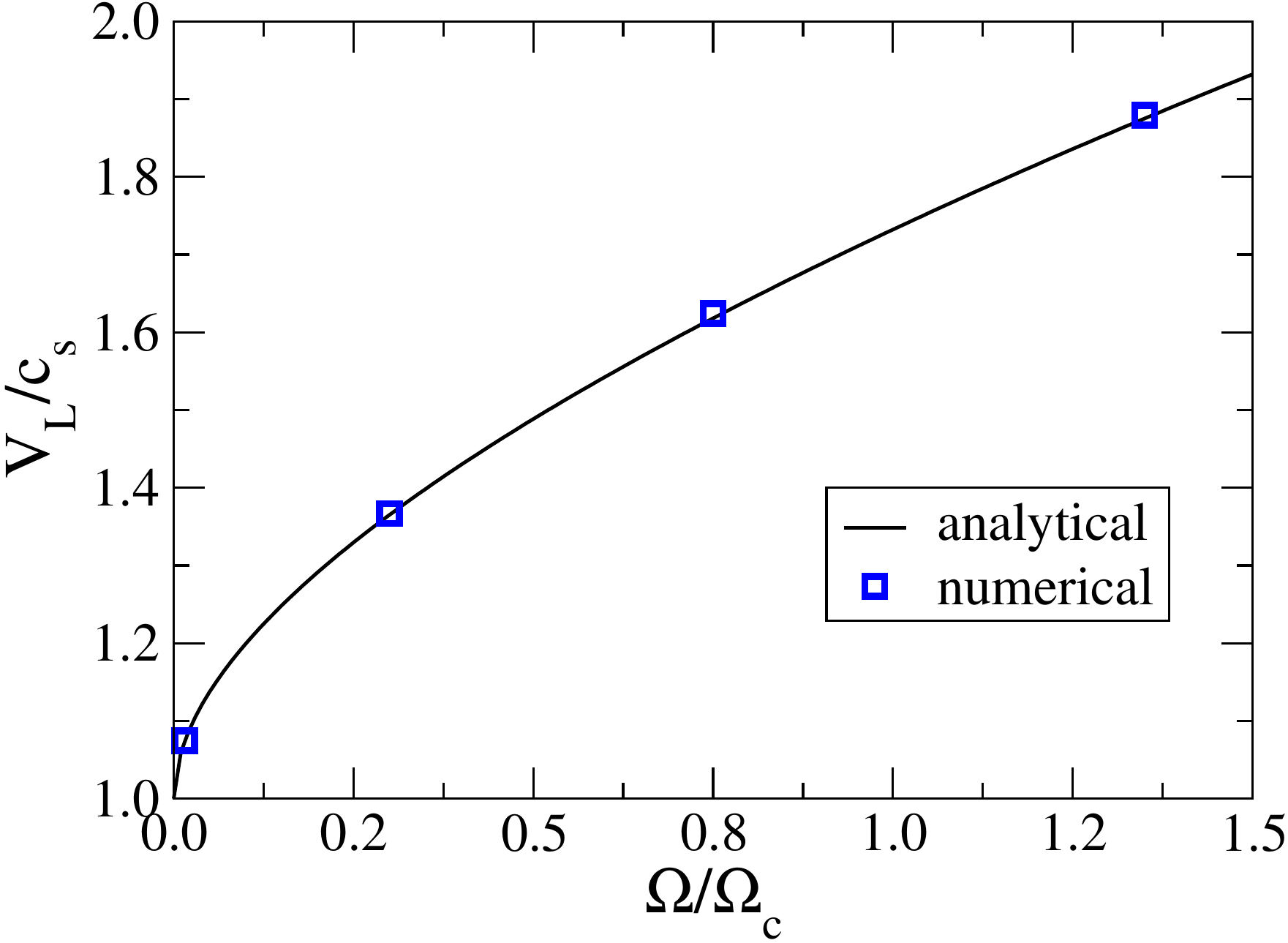}
\caption{Landau critical velocity for the disappearance of $0\protect\pi$
magnetic solitons as a function of the Rabi-coupling. The solid line shows the
analytic prediction and blue squares represent the numerical results 
for the points where the
energy of the $0\protect\pi$ solitons tends to 0.}
\label{fig:criticalU}
\end{figure}

The phase diagram in Fig.~\ref{fig:energy} shows that $0\pi $ magnetic
solitons will eventually disappear (i.e, the energy $E\rightarrow 0$) when
their velocity tends to a critical value. This critical velocity (hereafter
called Landau's critical velocity) is determined by Landau's criterion,
\begin{equation}
V_{\text{L}}=\text{min}_{p}\frac{\varepsilon _{\text{s}}(p)}{p},
\label{eq:criticalU}
\end{equation}%
associated with the emergence of an energetic instability in the dispersion
of the Bogoliubov spectrum 
\begin{equation}
\varepsilon _{\text{s}}\ =\sqrt{\left( \frac{\hbar ^{2}k^{2}}{2m}%
+\hbar\Omega \right) \left( \frac{\hbar ^{2}k^{2}}{2m}+\hbar\Omega +n\delta
g\right) }
\end{equation}%
of spin excitations in the presence of Rabi coupling~\cite%
{Goldstein1997,Tommasini2003,Abad2013}. Using Eq.~(\ref{eq:criticalU}) one finds the
result 
\begin{equation*}
\frac{V_{\text{L}}}{c_{\text{s}}}=\sqrt{1+\frac{2\hbar\Omega }{n\delta g}+%
\sqrt{\frac{2\hbar\Omega }{n\delta g}\left( \frac{2\hbar\Omega }{n\delta g}%
+2\right) }}
\end{equation*}%
for Landau's critical velocity which, in dimensionless form, reads 
\begin{equation}
U_{\text{L}}=\sqrt{1+\frac{2\omega _{\text{R}}}{3}+2\sqrt{\frac{\omega _{%
\text{R}}}{3}\left( 1+\frac{\omega _{\text{R}}}{3}\right) }}.
\end{equation}%
Figure~\ref{fig:criticalU} shows that the critical velocity extracted from the
phase diagram in Fig.~\ref{fig:energy} is in excellent agreement with the
above analytic prediction.

It is worth noting that when the velocity of the $0\pi$ soliton tends to
the Landau critical velocity, not only does its amplitude decreases, but also its
structure changes. The number of oscillations in the magnetization increases
and the soliton turns into a wide oscillating object in space (see Fig.~\ref%
{fig:typeIIcritical}). This fact is in accordance with the so-called theory
of soliton bifurcation discussed in Ref~\cite{KD11}. We leave
this for future investigation.

\begin{figure}[tbp]
\includegraphics[width=8.0cm]{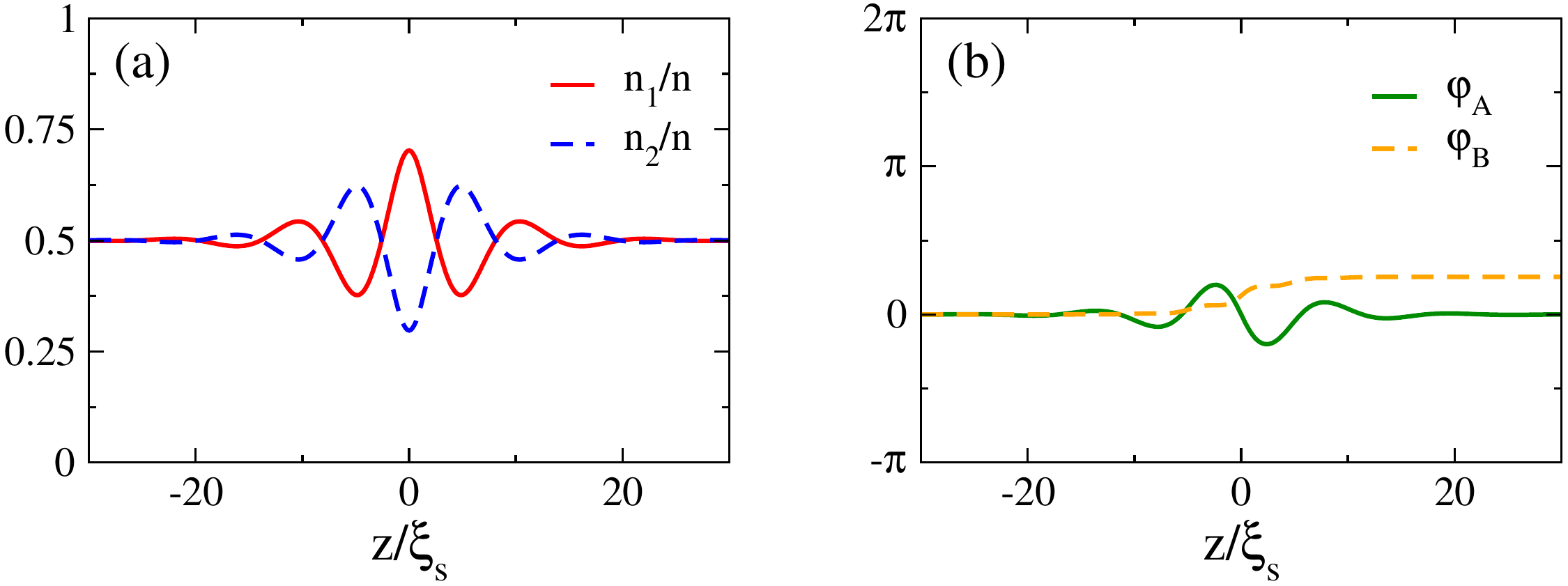}
\caption{Profiles of the $0\protect\pi$ magnetic soliton at velocity $V/c_%
\text{s}=-1.3$, $m_0=0.405$, and Rabi coupling $\protect\omega_\text{R}=0.3$.
(a) The solid red and dashed blue lines show the density distributions of the
two components, satisfying $(n_1+n_2)/n=1$. (b) The solid green and dashed orange
lines show the relative phase $\protect\varphi_A$ and total phase $%
\protect\varphi_B$ as a function of the coordinate. For $0\protect\pi$
solitons the asymptotic jump of $\protect\varphi_A$ is 0. Compared to
Fig.~\protect\ref{fig:typeII}, there are more oscillations because the
velocity of the soliton is close to the Landau critical velocity.}
\label{fig:typeIIcritical}
\end{figure}

\section{Dynamics in a 1D harmonic trap}

\label{sec:1dtrap} In the above sections, we focus on the exact solutions
for $2\pi$ and $0\pi$ solitons propagating in uniform matter, where their
shape and velocity remain unchanged during the motion. However, real
experiments are always implemented in trapped systems, where the density of
the condensate varies as a function of the position. The amplitude and velocity
of magnetic solitons are then expected to change in the trap.
In this section, we discuss the dynamics of magnetic solitons in a 1D
harmonic trapping potential $V_{\mathrm{ext}}(z)=m\omega _{\text{ho}%
}^{2}z^{2}/2$, with $\omega _{\text{ho}}$ as the trapping frequency.

If the
condensate size $L_{z}$ is large enough in comparison with $\xi _{\text{phase}}$, one
can solve this problem by using the conservation of the energy of the soliton and
the local density approximation, as has been performed for dark solitons 
in a single component BEC~\cite{Busch2000,Konotop2004}
and for magnetic solitons in the absence of Rabi coupling~\cite{Qu2016}.
Using results from Sec.~\ref{sec:energy4mag} in the dimensional form, the energy of the
soliton with its center at point $Z$ can be expressed as 
\begin{equation}
E(Z,V)=\frac{\hbar }{2}\sqrt{\frac{\delta g}{2m}}n^{3/2}(Z) \epsilon \left( 
\frac{\hbar \Omega }{n(Z)\delta g},
\frac{V}{\sqrt{n(Z)\delta g/(2m)}}
\right) 
\label{EZV}
\end{equation}%
where $n(Z)=n(Z=0)-V_{\mathrm{ext}}(Z)/g$ is the Thomas-Fermi equilibrium
density, $V(Z)=dZ/dt$ is the velocity of the soliton, and $\epsilon $ is a dimensionless function. Then the
energy conservation of a moving soliton can be written as $E(Z,V)=E(Z_{0},V_{0})$, where $Z_{0},V_{0}$ are the initial values of the position and velocity of the soliton,
allowing one to find $dZ/dt$ as a function of $Z$ and, consequently, to solve $Z(t)$ after a simple
integration.

However, the absence of an explicit analytical expression for the energy in
the presence of Rabi coupling makes this approach inconvenient and we therefore numerically solved
the dynamical problem with the help of the time-dependent coupled GPEs, corresponding to the
Lagrangian density, Eqs.~(\ref{eq:lagrangiandensity}) and (\ref{eq:energydensity}%
), exploiting in a more systematic way the main features anticipated Sec.~\ref{sec:intro}.

\begin{figure}[tbp]
\includegraphics[width=8.5cm]{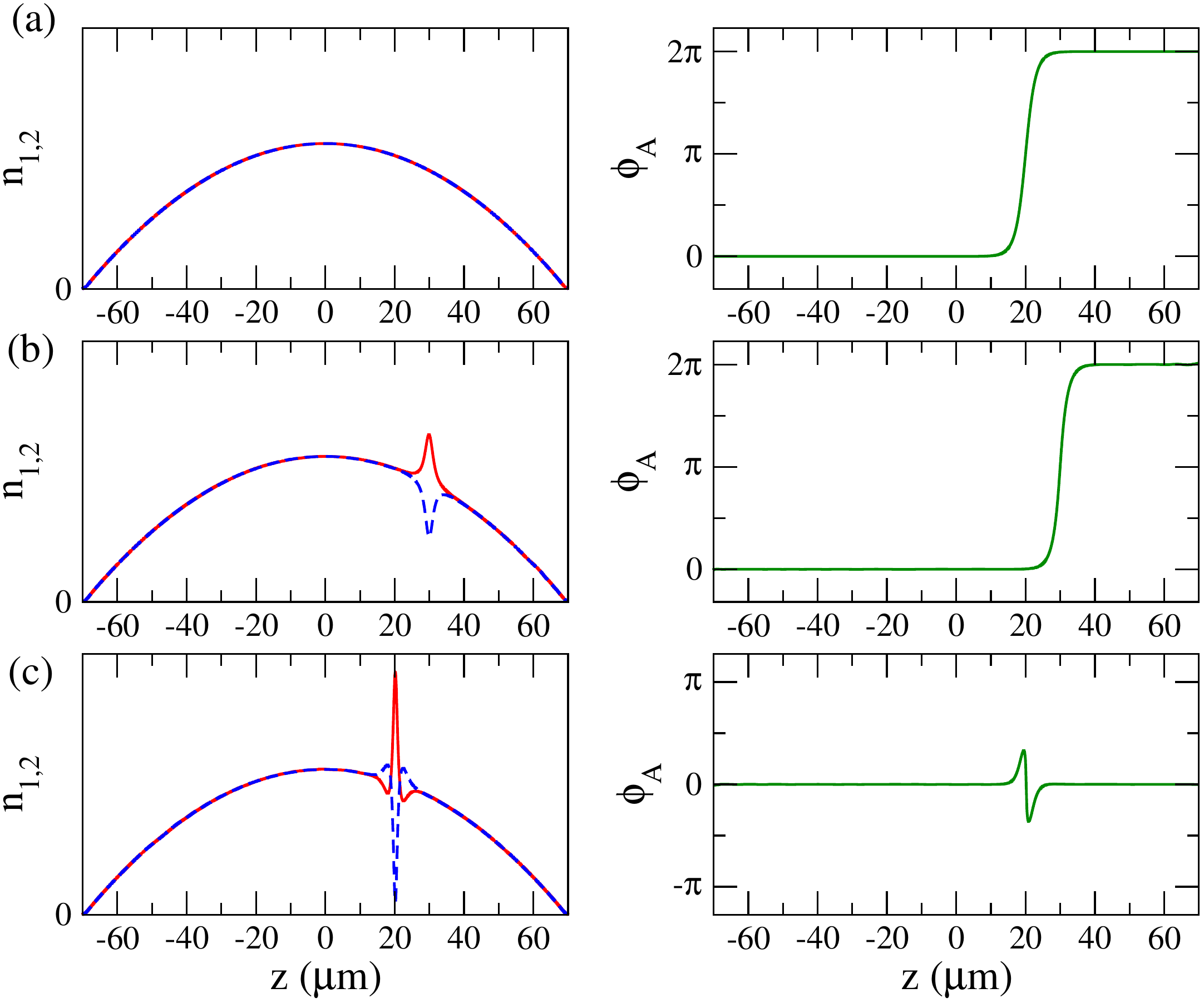}
\caption{Oscillation of magnetic solitons in a 1D harmonic trap. We imprint a Son-Stephanov domain wall
at the initial position $z_0=20\mu m$, the local Rabi coupling is given by 
$\protect\omega_\text{R}(z_0)=\Omega/\Omega_\text{c}(z_0)=0.22$. Evolutions of the densities and relative phase of the two components after
a holding time of (a) $\omega_\text{ho}t=0$ (b) $6.3$, and (c) $12.6$. }
\label{fig:1Dprofile}
\end{figure}

\begin{figure}[tbp]
\includegraphics[width=7.5cm]{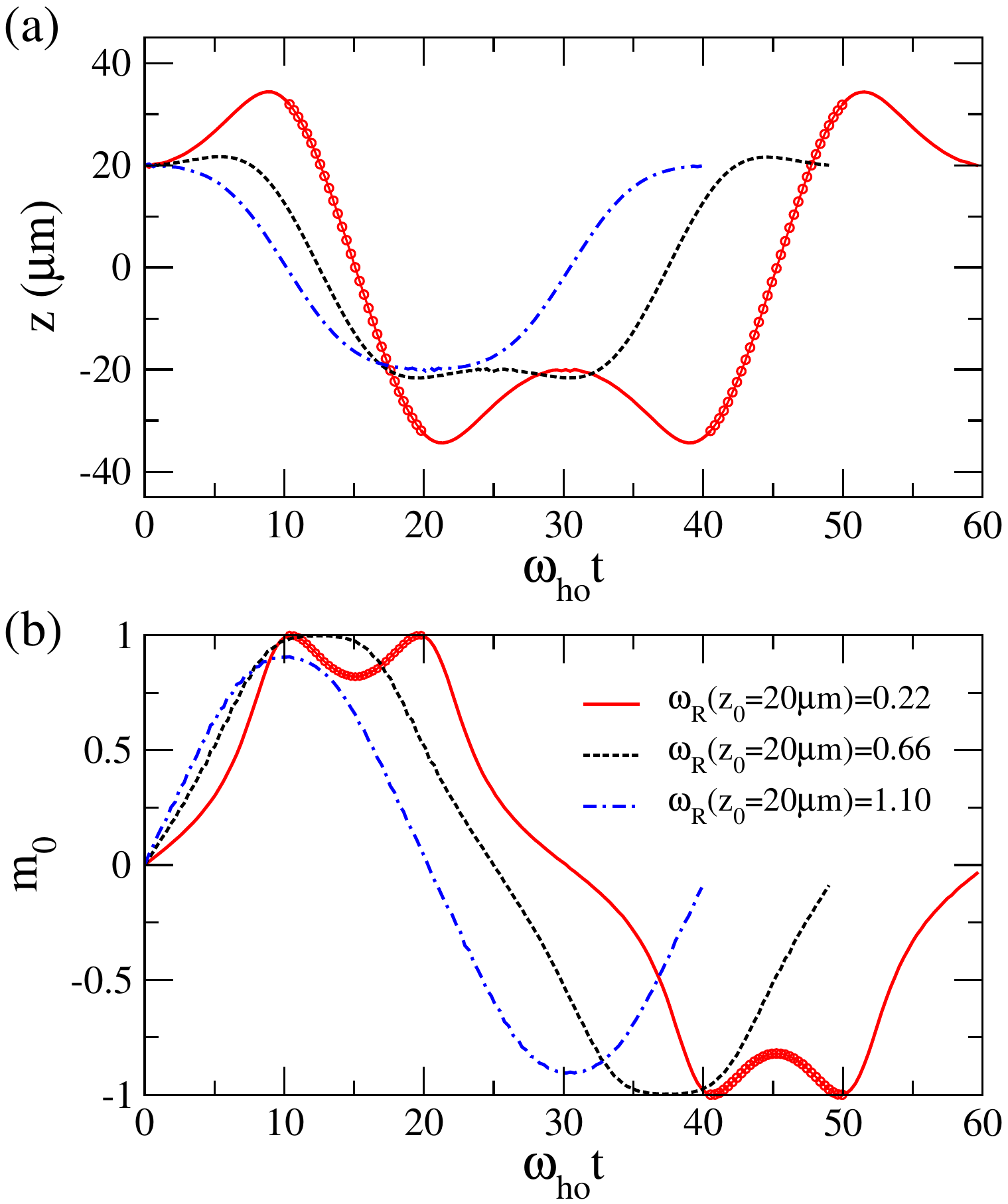}
\caption{(a) In-trap trajectories of magnetic solitons and (b) evolution of
the magnetization $m_0$ at the soliton center after imprinting a
Son-Stephanov domain wall at $z_0$ for different values of the local Rabi
coupling $\protect\omega_\text{R}(z_0)=\Omega/\Omega_\text{c}(z_0)=0.22$
(solid red line), $0.66$ (dashed black line), $1.1$ (dash-dotted blue line). Lines without circles indicate $2\protect\pi$
solitons; lines with circles, $0\protect\pi$
solitons. Spin-interaction parameters have been chosen the same as in $%
^{23}$Na, where $\protect\delta g=0.07g$ and the Thomas-Fermi radius is $R_%
\text{TF}=70\protect\mu m$. All curves presented here correspond to the
time interval of an oscillation period.}
\label{fig:trajectory1}
\end{figure}

\begin{figure}[tbp]
\includegraphics[width=7.5cm]{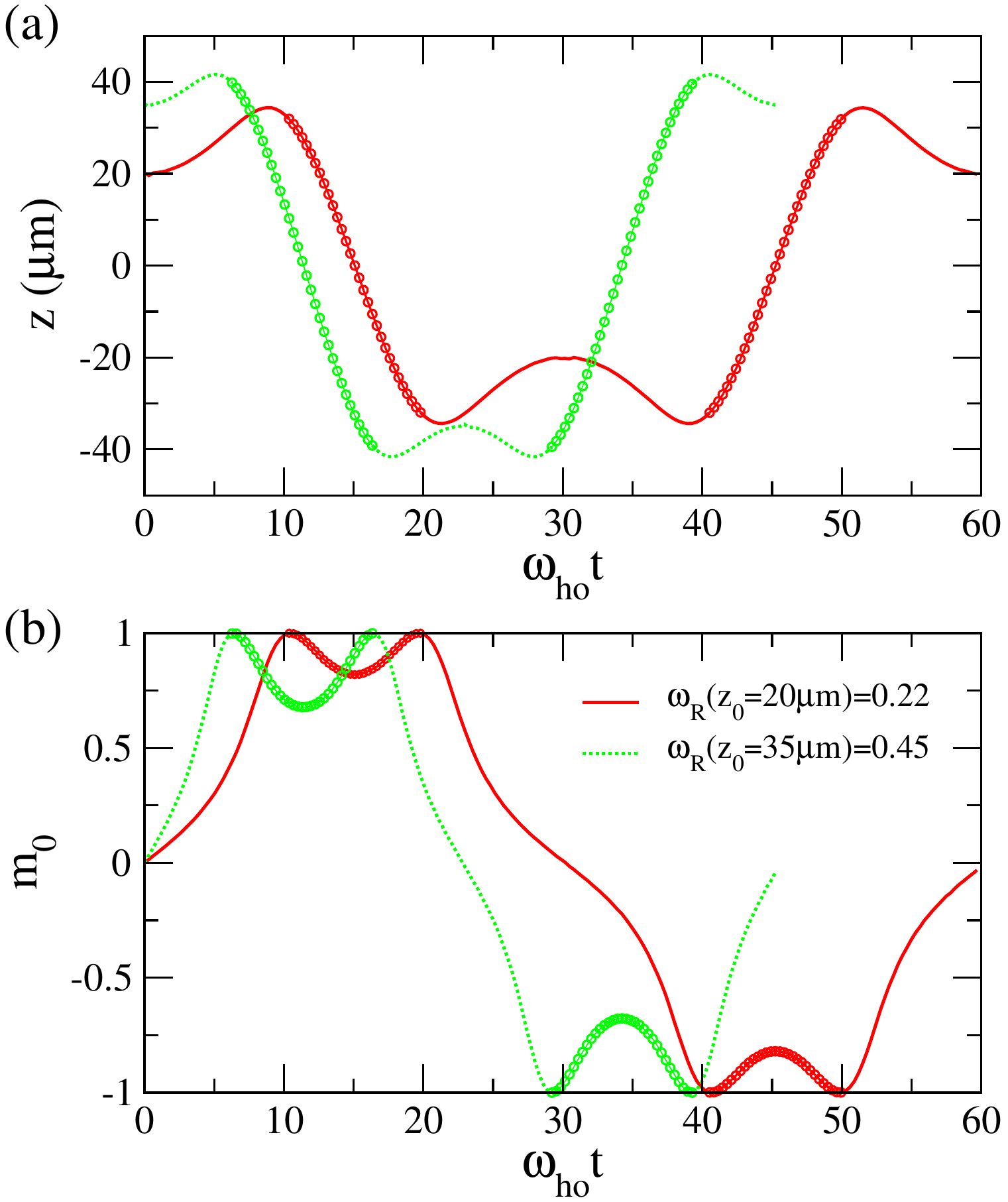}
\caption{(a) In-trap trajectories of magnetic solitons and (b)
evolution of the magnetization $m_0$ at the soliton center after imprinting
a Son-Stephanov domain wall at different initial positions $z_0=20\protect%
\mu m$ (solid red line) and $35\protect\mu m$ (dotted blue line). The corresponding local
dimensionless Rabi couplings are $\protect\omega_\text{R}(z_0)=0.22$ (red)
and $0.45$ (blue), respectively. Lines without circles indicate $2\protect\pi$
solitons; lines with circles, $0\protect\pi$
solitons. Spin-interaction parameters have been chosen the same as $%
^{23} $Na with $\protect\delta g=0.07g$ and the Thomas-Fermi radius is $R_%
\text{TF}=70\protect\mu m$.}
\label{fig:trajectory2}
\end{figure}

To investigate the oscillation dynamics of magnetic solitons in a 1D harmonic trap, we imprint a Son-Stephanov domain wall
at the initial position $z_0$ and then observe the evolution of the densities and relative phase of the two components after
a holding time. An example is shown in Fig.~\ref{fig:1Dprofile}. The domain wall was first imprinted at $z_0=20\mu m$. It moves to the right edge of the trap and becomes a $2\pi$ magnetic soliton. After a longer evolution time, it returns and moves toward the trap center, and evolves into a $0\pi$ magnetic soliton.

In Fig.~\ref{fig:trajectory1} and Fig.~\ref{fig:trajectory2}, we track the
trajectories of the magnetic solitons after the imprint of a Son-Stephanov
domain wall at $z_0$ for a complete oscillation period. Note that the
external Rabi coupling $\Omega$ is a constant for each simulation. However,
in the presence of harmonic trap, the density varies and the local
dimensionless Rabi coupling $\omega_\text{R}(z)=\Omega/\Omega_{\text{c}}(z)$
is also position dependent, its value being minimum at the trap center and
very large near the border of the atomic cloud.

In Fig.~\ref{fig:trajectory1}, we study the oscillation dynamics of the
magnetic solitons for different values of Rabi-coupling $\Omega$ [and thus
different values of $\omega_\text{R}(z_0)$] after imprinting a domain wall
at the same initial position, $z_0=20\mu m$. With the increase in $\omega_%
\text{R}(z_0)$, the region exhibiting $0\pi$ solitons shrinks and eventually
disappears. Furthermore, the anharmonic oscillations in the presence of both 
$2\pi$ and $0\pi$ magnetic solitons (see red curve) tend to become
harmonic when $0\pi$ solitons are no longer produced during the oscillation.
The black curve corresponds to the case where $0\pi$ solitons no longer
emerge during the oscillation. The blue curve reveals that, a $2\pi$
soliton, imprinted in a 1D harmonically trapped condensate under the
condition $\omega_\text{R}(z_0)>\omega_\text{R}^\text{c}(z_0)$ of the Rabi
coupling, despite its instability \cite{Son2002}, exhibits a regular
oscillation moving first towards the center of the trap because of its
negative effective mass.

The analysis of Fig.~\ref{fig:trajectory1} shows that, in order to observe
the emergence of both $2\pi$ and $0\pi$ solitons during the oscillation, the
local Rabi coupling at the initial position $z_0$ should be significantly
smaller than $\omega_\text{R}^\text{c}(z_0)$.

In Fig.~\ref{fig:trajectory2}, we study the oscillation dynamics of the
magnetic solitons for different initial positions of the phase imprinting.
For larger $z_0$, the density of the condensate is lower and thus $\omega_%
\text{R}(z_0)$ is larger. In this case, the $2\pi$ magnetic soliton reaches
the turning point faster and the $0\pi$ soliton appears earlier.

Finally, we remark that although our theory has been based on the assumption
of condition~(\ref{eq:condition1}), a similar phenomenon also occurs for larger
values of $\delta g$ where the total density exhibits a dark soliton. To
demonstrate this, we relax condition~(\ref{eq:condition1}) and
present the simulation dynamics in the presence of larger $\delta g$ in the
following investigation of the role of transverse confinement.

\section{Role of the transverse confinement}

\label{sec:2dtrap}

\begin{figure}[!t]
\includegraphics[width = 0.75\columnwidth]{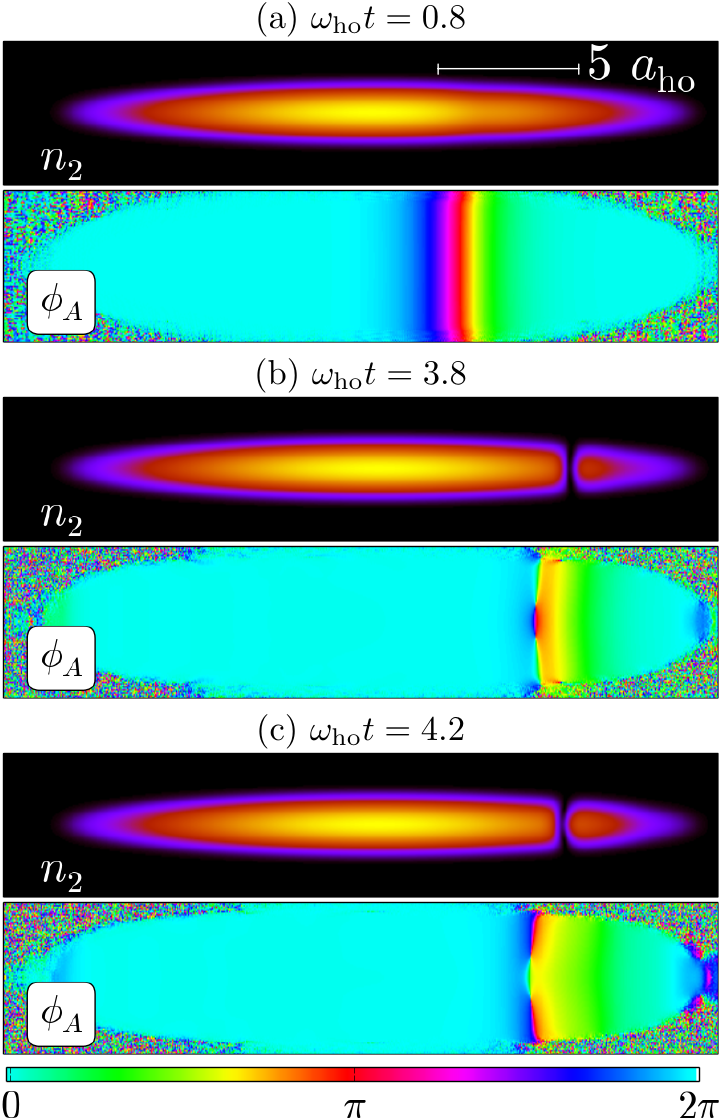}
\caption{Evolution of the magnetic solitons in an elongated harmonic trap
(aspect ratio $=10$) after imprinting a Son-Stephanov domain wall. For each
time instant we show the density of the second component $n_2$ (the
brighter the color the higher the density) in the upper
panel and the relative phase $\protect\varphi_A$ (the color or grey
scale changes continuously from $0$ to $2 \pi$) in the lower panel:
(a) $\omega_{\rm ho} t = 0.8$, (b) $\omega_{\rm ho} t = 3.8$, (c)
$\omega_{\rm ho} t = 4.2$. Rabi coupling $\Omega = 0.5\,
\omega_\text{ho}$ and interaction $\delta g = 0.4\, g$. In the top
panel we also show a length scale corresponding to $5\, a_{\rm ho}$. }
\label{fig:2D.elongated}
\end{figure}

\begin{figure}[b!]
\includegraphics[width = 1\columnwidth]{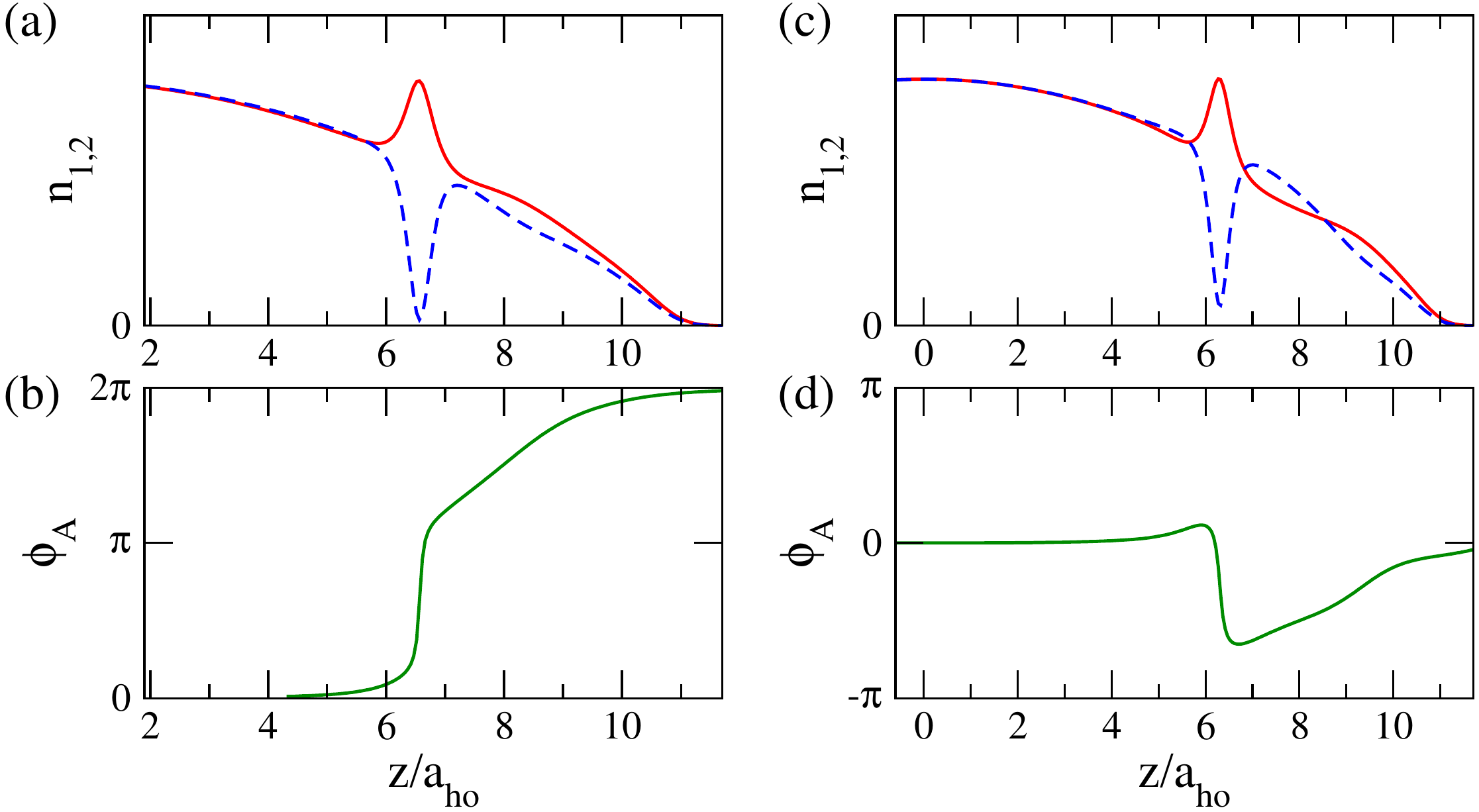}
\caption{Densities of the two spin components and their relative phase $%
\protect\varphi_A$ calculated on a cut along the longer axis of the
configurations presented in Fig.~\protect\ref{fig:2D.elongated}. Soliton
(left) before ($\protect\omega_\text{ho} t = 3.8$) and (right) after ($\protect\omega_\text{%
ho} t = 4.2$) reflection. We can see that the configuration before the
reflection corresponds to a $2\protect\pi$ soliton and the configuration
after the reflection corresponds to a $0\protect\pi$ soliton.}
\label{fig:2D.elongated_cut}
\end{figure}

In this section we generalize our results to two-dimensional (2D)
configurations. In 2D or higher dimension, solitons with a negative effective mass are expected to be unstable due to snake instability. However, for elongated geometry, it is still possible to observe persistent oscillations of the solitons. The size of the transverse confinement ensuring stability of the moving domain wall or magnetic soliton can be roughly estimated as $R_{\perp} < \xi_{\rm phase}$. For parameters used in our calculations, $\Omega = 0.5\, \omega_\text{ho}$, $\mu \approx 50\, \hbar \omega_\text{ho}$, and $\delta g = 0.4\, g$, this gives an aspect ratio $\omega_\perp / \omega_{\rm ho} > 2 \sqrt{\mu \Omega / (\hbar \omega_{\rm ho}^2)} = 10$. We first consider the case of an elongated harmonic trap, with aspect ratio $\omega_\perp /\omega_\text{ho}= 10$, where $\omega_%
\text{ho}$($\omega_\perp$) is the harmonic trapping frequency along the
longitudinal (transverse) direction, in order to understand how the 1D
solutions behave in this elongated geometry. We expect that this elongated
geometry will share many features with one dimension. Indeed, the domain wall characterized
by a $2\pi$ relative phase jump, which was initially imprinted along the
weak axis of the trap and displaced from the center by a small fraction of
the Thomas-Fermi radius, begins to travel along the weak axis towards the
closer edge of the cigar could. When the $2\pi$ soliton moves to the turning
point, it develops a density polarization and induces two vortices at its
ends (see Fig.~\ref{fig:2D.elongated}). Then, it moves back towards the
center of the trap as predicted for the 1D solution, but now we observe
that the soliton is fragmented into two pieces and no longer extends
through the whole transverse dimension [see Fig.~\ref{fig:2D.elongated}(c)].
As discussed in~\cite{Tylutki2016} and ~\cite{Calderaro2016} the end of a finite domain
wall is always associated with the existence of a vortex in one of the two
spin components, ensuring the proper behavior of the phase around the end
point. In the region between the vortices we have a polarized density, which suggests that our solution matches the $0\pi$ magnetic soliton obtained
in the 1D configuration. The $0\pi$ soliton continues to move and it survives for a long time while oscillating and repeatedly transforming to $2\pi$ solitons in the harmonic trap. In Fig.~\ref{fig:2D.elongated_cut} we show the cut
of the density and phase of the gas along the weak confinement axis before
and after the reflection. We can recognize the same structure as in Fig.~\ref%
{fig:typeI} and Fig.~\ref{fig:typeII} for $2\pi$ and $0\pi$ solitons,
respectively.

\begin{figure}[tb]
\includegraphics[width = 0.8\columnwidth]{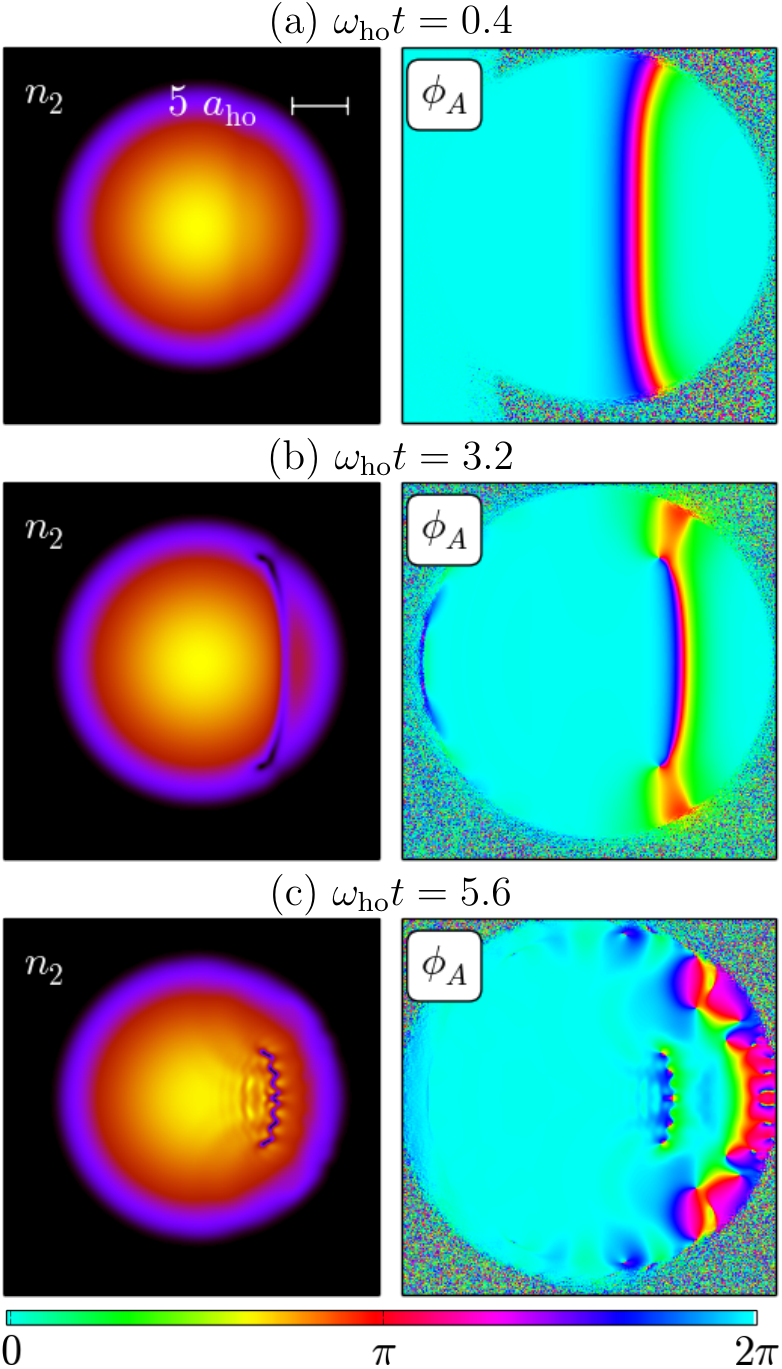}
\caption{Evolution of an imprinted Son-Stephanov domain wall in a spherical
harmonic trap. For each time instant we show the density of the second
component $n_2$ (the brighter the color the higher the density) in the
upper panel and the relative phase $\protect\varphi_A$ (the color or
grey scale changes continuously from $0$ to $2 \pi$) in the lower
panel: (a) $\omega_{\rm ho} t = 0$, (b) $\omega_{\rm ho} t = 3.2$, (c)
$\omega_{\rm ho} t = 5.6$. Rabi coupling $\Omega = 0.5\,
\omega_\text{ho}$ and interaction $\delta g = 0.4\, g$. In the top
left panel we also show a length scale corresponding to $5\, a_{\rm
ho}$.  }
\label{fig:2D.isotropic}
\end{figure}

For fully 2D configurations, the 1D dynamics of magnetic solitons discussed
in the previous sections no longer applies, and the soliton cannot
oscillate indefinitely due to the snake instability. To demonstrate this, we have repeated the numerical
simulation for an isotropic harmonic potential, where the 2D physics should
be fully manifested. We have assumed $\Omega = 0.5 \omega_\text{ho}$ and $%
\delta g = 0.4 g$. Initially the $2\pi$ soliton travels to the edge of the
trap, similarly to the case of an elongated trap. However, the soliton starts to bend, and the vortices
generated near the end of the soliton become detached from the cloud boundary.
Furthermore, the reflection is associated with the production of multiple
vortices along what was formerly a single soliton (see Fig.~\ref%
{fig:2D.isotropic}). These vortices travel back towards the center together,
but soon the dynamics becomes very complicated. The excess energy is
converted into phononic excitations, and the soliton is lost.
\\

\section{Discussion}

\label{sec:discuss} We have investigated the main features of moving
magnetic solitons in Rabi-coupled binary Bose-Einstein condensates. Two
types of magnetic solitons have been identified and characterized: (i) $2\pi$
solitons, which are connected to the unmagnetized static Son-Stephanov
domain wall and exhibit a $2\pi$ relative phase jump; (ii) $0\pi$ solitons,
which are connected to $2\pi$ solitons at a critical velocity, where the
density of one component vanishes, and which do not exhibit a net jump of
the relative phase. The complete phase diagram, the energy, and the magnetic
properties of these solitons are obtained in a uniform matter, and their
dynamical evolution is calculated in a 1D and 2D harmonic trap. A peculiar
feature emerging from our calculations is that $2\pi$ solitons evolve into $%
0\pi$ solitons (and vice versa) during their oscillatory motion in a
harmonic trap.

We expect that these novel examples of solitons can be observed experimentally in
the near-future. To observe them in ultracold atoms one can, for example,
use a mixture of the $|F=1,m_F=+1\rangle$ and $|F=1,m_F=-1\rangle$ hyperfine
components of the $3^2S_{1/2}$ states of sodium, where $\delta g/g\approx
0.07$~\cite{Knoop2011} and the exact miscibility of the atomic states can be
reached~\cite{Bienaime2016}. For typical experimental parameters, the
chemical potential is $\mu\sim h\times 10^4$Hz, and thus the critical Rabi
coupling is estimated as $\Omega_c=n\delta g/3=0.023ng=0.023\mu=h\times 230$%
Hz. Therefore, a weak Rabi coupling (of the order of $\sim 100$Hz) is required
to observe these magnetic solitons, a condition which can be achieved with
current experimental techniques.

Although our discussion of magnetic solitons has been focused on the context
of binary Bose-Einstein condensates, similar physics can be easily
generalized to and investigated in other physical systems which are governed
by coupled GPEs, such as fiber optics~\cite{Akhmediev1993} and
exciton-polaritons~\cite{Solnyshkov2012}.

\begin{acknowledgements}
We would like to thank Gabriele Ferrari, Anatoly Kamchatnov and William D. Phillips for useful discussions. This work was supported by the QUIC grant of the Horizon2020 FET program and by Provincia Autonoma di Trento. M.T. was partially supported by the PL-Grid infrastructure.
\end{acknowledgements}


\begin{thebibliography}{99}
\bibitem{footnote0} Note that in the presence of Rabi coupling, the
stability of the mixture can be even ensured for values of $\delta g<0$,
provided the condition $\hbar\Omega>-n\delta g$ is satisfied. (see Chap. 21 of Ref.~\cite{Pitaevskii2016})
\bibitem{footnote} Generally speaking, the variation should be performed
before substituting $\partial _{\zeta }\varphi _{B}$ into $\tilde{\mathcal{L}%
}.$ However, in our case it is possible to first implement this substitution
due to the solitonic boundary conditions imposed on $\varphi _{B}$. Indeed,
the variation should be performed at constant $\varphi _{B}$. The
substitution introduces additional dependence on $\varphi _{A}$ and $\theta $%
. However, this dependence does not contribute to variations, because due to
Eq.~(\ref{varfB}) and Eq.~(\ref{dfBdz}) we get $\partial \tilde{L}/\partial
\left( \partial _{\zeta }\varphi _{B}\right) =0$. In more general cases one
should performe variation of so-called Rauth function instead of function $%
\tilde{\mathcal{L}}$. (See, for example, Ref~\cite{LL1}, Sec. 41 or Ref~\cite%
{G00}, Sec. 8.3.)
\bibitem{footnote2} It is worth to note that to obtain a correct solution,
the parameter $\theta_0$ should be tuned with a very high precision.
\end{thebibliography}
\end{document}